\documentclass[fleqn,usenatbib]{mnras}
\usepackage{newtxtext,newtxmath}

\usepackage[T1]{fontenc}
\usepackage{ae,aecompl}
\usepackage{graphicx}   
\usepackage{amsmath}    
\usepackage{amssymb}    
\usepackage{booktabs}
\usepackage{multirow}
\usepackage[detect-all,version-1-compatibility]{siunitx}

\title[Scalable End-to-End Recurrent Neural Network]{Scalable End-to-End Recurrent Neural Network for Variable Star Classification}

\author[Becker et al.]{
I. Becker$^{1,2}$\thanks{E-mail: iebecker@uc.cl},
K. Pichara$^{1,2,4}$\thanks{E-mail: kpb@ing.puc.cl},
M. Catelan $^{2,3,4}$\thanks{E-mail: mcatelan@astro.puc.cl},
P. Protopapas$^{5}$\thanks{E-mail: avlos@seas.harvard.edu},
C. Aguirre$^{1}$,
F. Nikzat$^{2,3}$
\\
\\
$^{1}$Pontificia Universidad Cat\'{o}lica de Chile, Computer Science Department, Santiago, Chile\\
$^{2}$Millennium Institute of Astrophysics, Santiago, Chile\\
$^{3}$Pontificia Universidad Cat\'olica de Chile, Facultad de F\'{i}sica, Instituto de Astrof\'\i sica, Av. Vicu\~{n}a Mackenna 4860, 7820436 Macul, Santiago, Chile\\
$^{3}$Pontificia Universidad Cat\'olica de Chile, Centro de Astro-Ingenier\'{i}a, Av. Vicu\~{n}a Mackenna 4860, 7820436 Macul, Santiago, Chile\\
$^{5}$Institute for Applied Computational Science, Harvard University, Cambridge, MA, USA
}

\date{Accepted XXX. Received YYY; in original form ZZZ}

\pubyear{2020}

\begin{document}
\label{firstpage}
\pagerange{\pageref{firstpage}--\pageref{lastpage}}
\maketitle

\begin{abstract}
During the last decade, considerable effort has been made to perform automatic classification of variable stars using machine learning techniques. Traditionally, light curves are represented as a vector of descriptors or features used as input for many algorithms. Some features are computationally expensive, cannot be updated quickly and hence for large datasets such as the LSST cannot be applied. Previous work has been done to develop alternative unsupervised feature extraction algorithms for light curves, but the cost of doing so still remains high. 
In this work, we propose an end-to-end algorithm that automatically learns the representation of light curves that allows an accurate automatic classification. We study a series of deep learning architectures based on Recurrent Neural Networks and test them in automated classification scenarios. Our method uses minimal data preprocessing, can be updated with a low computational cost for new observations and light curves, and can scale up to massive datasets. We transform each light curve into an input matrix representation whose elements are the differences in time and magnitude, and the outputs are classification probabilities. We test our method in three surveys: OGLE-III, Gaia and WISE. We obtain accuracies of about $95\%$ in the main classes and $75\%$ in the majority of subclasses. We compare our results with the Random Forest classifier and obtain competitive accuracies while being faster and scalable. The analysis shows that the computational complexity of our approach grows up linearly with the light curve size, while the traditional approach cost grows as $N\log{(N)}$.  
\end{abstract}



\section{Introduction}\label{section:Introduction}
Variable stars comprise a significant fraction of all stars. They have been essential to build the distance ladder, identify dwarf galaxies, and to enable us to understand the formation history of our galaxy, among others \citep[see, e.g.,][for a broad overview of stellar variability and applications]{Catelan2015}. Classification of variable stars is one of the critical processes in a time-domain survey since it enables us to extract the most from the data. 

Astronomers rely on automatic classification algorithms since the amount of data is impractical to be manually analyzed. Even for experienced astronomers, light curve classification presents a challenge since they are measured at irregular intervals, often are noise dominated and do not have the same number of measurements among them. A preprocessing step is imperative to classify variable stars. Usually, astronomers compute statistical descriptors or features \citep{Bloom2012, Nun2015}. Features condense information of each light curve into a vector of finite length. Some of these descriptors can be computationally expensive when a new light curve or new observations for an existing light curve are available since some of them need to be recomputed. 

In their seminal work, \citet{Debosscher2007} constructed a set of 28 features to describe light curves. They are based on analytical fits, and periods obtained using the Lomb-Scargle periodogram \citep{Lomb1976, Scargle1982}. Selected features included light curve mean magnitudes, standard deviation, median and amplitude, among others. On the same line, the work of \citet{Kim2011} aimed to detect Quasi-Stellar Objects (QSOs) in the MACHO \citep{Alcock2000} dataset. They found that the Random Forest classifier (RF; \citealt{Breiman2001}) outperforms Support Vector Machine (SVM; \citealt{Cortes1995}) using 11 features. The same year \citet{Richards2011} obtained the same conclusions using 53 features. Recent works still apply the same techniques to find specific classes of variability \citep{Elorrieta2016, Gran2016}.

\citet{Pichara2012}, working with the EROS database \citep{Beaulieu1995}, included autoregressive features which prove valuable for QSO classification. The model was enhanced later by \citet{Pichara2013}, who used graphical models to fill missing data while keeping the computational cost the same. \citet{Kim2014} used 22 features for classifying classes and subclasses of variable stars using an RF, which defined that algorithm as the best performer in classification.

The RF algorithm has an integrated feature selection and evaluation mechanism. It gives an enormous degree of flexibility while maintaining robust results. Thus, increasing the number of features only increases its performance. That is the reason why more features kept being implemented to be used with this algorithm. 

A classifier's performance depends on the quality and quantity of the information they are trained on. Surveys have been generating data for decades, but the number of objects, data volume, and generation rate are increasing exponentially \citep{Huijse2014, Garofalo2016}. This scenario will become even more challenging in 2022 when the Large Synoptic Survey Telescope (LSST, \citealt{Abell2009}) starts science operations. LSST will produce around 15\,TB of data per night which must be processed in real-time to extract the best science from it.

As such, the classification is not the only problem that needs to be tackled. Other problems that emerge from working with data from different surveys. They have not only different science goals but also have different cadence, optics, detectors, reduction pipelines, among others. These differences produce various biases and systematic errors that need to be treated independently for each dataset, which further increases the difficulty of the machine learning approach.

Nonetheless, there have been numerous efforts to publish already trained tools to ease the process for non-experts. The work of \citet{Nun2015} implemented the automatic feature extraction package FATS (Feature Analysis for Time Series), while \citet{Kim2016} published a feature extraction package and a trained classifier. However, it should be noted that a trained classifier might not have the same performance for different surveys and cannot always be re-trained to suit specific needs. 

The technological challenges imposed by the LSST are plenty \citep{Ivezic2016}. One of those challenges is the alerts and broker systems. It will need to classify and emit alerts for 10 million events every night and keep track of 37 billion followed objects. Since the telescope has to fulfil all of its science goals, the cadence will not be optimized to study variable or transient objects and is still a topic of discussion inside the LSST community \citep{Marshall2017}. 

Training sets are usually are obtained from samples with good quality photometry. They might be biased towards brighter objects, not being entirely representative of the whole population and survey characteristics. The LSST will make discoveries at magnitudes at which no survey has observed before, making available datasets unfit, both in-depth and completeness. 

Features obtained from biased training sets might not work properly on the whole dataset. Furthermore, their design is time-consuming and usually only considers known behaviours. This excludes unknown objects, which further increases the biases. 

In recent years, unsupervised methods have been developed to create features without the direct intervention of a scientist (\citealt{Mackenzie2016, Gieseke2017, Valenzuela2017}). They achieve competitive results against human-designed features. Their advantage is that those features are optimized for the specific survey and have less bias than the human-crafted counterparts. In the novel work of \citet{Mackenzie2016}, they took a different approach, developing an unsupervised feature learning algorithm for variable stars. It uses a sliding window to sample fragments of a light curve. From them, it creates a dictionary of the most relevant fragments and uses them to make a new representation of the light curves. With this information, they trained an SVM classifier obtaining similar and even better results than the RF while reducing the computational cost significantly. 

While already classified variable stars might not change their behaviour, their representation does, depending on the survey they were observed. To avoid the re-computation of features, \citet{Benavente2017automatic} developed a method which transforms them directly from one survey to another. 
\citet{Pichara2016} proposed a meta-model which integrates different trained classifiers into a single framework, to avoid the re-training of models from scratch. It takes into account the context, representation of the data and computational complexity of each classifier. \citet{Valenzuela2017} proposed a model to perform unsupervised automatic classification of variable stars. The authors used an unsupervised feature extraction scheme and performed classification using a query-based method in tandem with a dedicated data structure.

Less supervised methods have proven successful in many areas. In the past years, Artificial Neural Networks (ANNs; see \citealt{Lecun2015} for a review) have shown its generalizing potential and ability to leverage vast amounts of data. These algorithms can create representations of the data using a sequence of linear and non-linear functions. Convolutional Neural Networks (CNNs, \citealt{LeCun1998convolutional}) are a type of ANN that is designed to capture local and global patterns in the input data. 

Only in the last years, methods involving ANNs have been applied to astronomy. \citet{Baglin2002} used a neural network to classify microlensing events. \citet{Dieleman2015} won the Galaxy Zoo Challenge by applying convolutional neural networks to galaxy morphology classification. \citet{Cabrera2017} used the same ideas to classify transients in the HiTS survey, surpassing results over an RF. \citet{Aguirre2018} developed a convolutional neural network to classify variable stars from multiple catalogues in a single model. This architecture removes the need for extracting features or separating datasets, since the network can detect common patterns, without needing the band information. \citet{Shallue2018} used them to detect extrasolar planet transits.

Recurrent Neural Networks (RNNs; \citealt{Lipton2015} for a review) are a family of architectures dedicated to sequential data. In the same way as crafted features, these networks can encode information in a fixed-length vector. These features are less biased as they are extracted from the data itself and not designed by scientists.

RNNs have been applied to time series classification. \citet{Charnock2017} applied an RNN framework to classify supernovae, obtaining competitive results against other classifiers. More recently, \citet{Naul2018recurrent} applied an encoder RNN as a feature extraction method and compared its performance against FATS \citep{Nun2015} using an RF. 

\citet{Charnock2017} proposed a bi-directional RNN to classify multi-band time series of supernovae. Their work yielded competitive results, but the size of the training set hindered their analysis. \citet{Naul2018recurrent} utilized a similar approach, but used a bidirectional RNN to encode and extract information as a feature extraction step, and then applied an RF classifier. Their features were comparable to the FATS features while being able to scale to bigger datasets.

New surveys impose strict restrictions on classifiers, as they need to process an unprecedented amount of information. As such, any algorithm working on such datasets must be efficient, in the dimension of the representation, the computational cost to obtain and update such representation. A smaller representation also optimizes the network resources, as less information needs to be transmitted to perform the classification of an object.

In this work, we propose a neural network that can leverage the vast amount of data available from surveys, while reducing the preprocessing needed to perform automatic classification of variable stars. Our approach does not compute features, can scale to vast amounts of data and can be updated with new information if so required, without the need for re-training the model.
We present our experimental analysis using three datasets: OGLE-III \citep{Udalski2004}, Gaia \citep{prusti2016gaia}, and WISE \citep{Wright2010}. These surveys were conducted using different photometric bands, and also differ in the number of observations and noise characteristics. We compare our results with the Random Forest classifier, which to date is the best performer \citep{Richards2011, Dubath2011, Long2012, Bloom2012, Gieseke2017} and widely used. 
 
This paper is organized as follows. 
In section \ref{section:Background_Theory} we introduce the relevant background theory on RNNs. Section \ref{section:Method_Description} presents the methodology used to process the data. Section \ref{section:Datasets} presents the datasets used , and Section \ref{section:Results} presents the results of our experiments. Finally, in section \ref{section:Conclusion}, we present our main conclusions and future work.

\section{Recurrent Neural Networks} \label{section:Background_Theory}
ANNs usually impose restrictions on the shape of its inputs, as all must be of the same size. Light curves, on the other hand, have a different number of observations measured at different times. A dedicated architecture must be used to avoid the extraction of features from the time series. RNNs are a group of architectures designed to process sequential data, such as time series, and does not need to compute features as the RF does.

For each element of a sequence $\boldsymbol{\mathit{x}}_\text{t}$ of $T$ observations, the network computes a vector $\boldsymbol{\mathit{h}}_\text{t}$ called hidden state. This vector encodes the information of the previous elements of the sequence up to that point. 

To compute it, the network operates the previous hidden state $\boldsymbol{\mathit{h}}_\text{t-1}$ and the data of the current time step $\boldsymbol{\mathit{x}}_\text{t}$, as described in Equation~\ref{eq:vanila_rnn},
\begin{equation}
 \boldsymbol{\mathit{h}}_\text{t} = f\left (\mathbf{W}\boldsymbol{\mathit{x}}_\text{t} + \mathbf{U} \boldsymbol{\mathit{h}}_\text{t-1} + \boldsymbol{\mathit{b}} \right ) \label{eq:vanila_rnn}.
\end{equation}

In this scenario, $\mathbf{W}$ matrix controls how much information of the current time will be used to compute $\boldsymbol{\mathit{h}}$; $\mathbf{U}$ matrix controls the information from the previous hidden state that will be used to update the hidden state; $\boldsymbol{\mathit{b}}$ is a bias term; and $f$ is a non-linear function, such as the sigmoid function, applied element-wise, which projects the values into a fixed range. 

 The above operations are grouped into a function called cell, which receives $\boldsymbol{\mathit{h}}_\text{t-1}$ and $\boldsymbol{\mathit{x}}_\text{t}$ and outputs $\boldsymbol{\mathit{h}}_\text{t}$. 
\begin{figure}
\centering
 \includegraphics[width=0.45\textwidth]{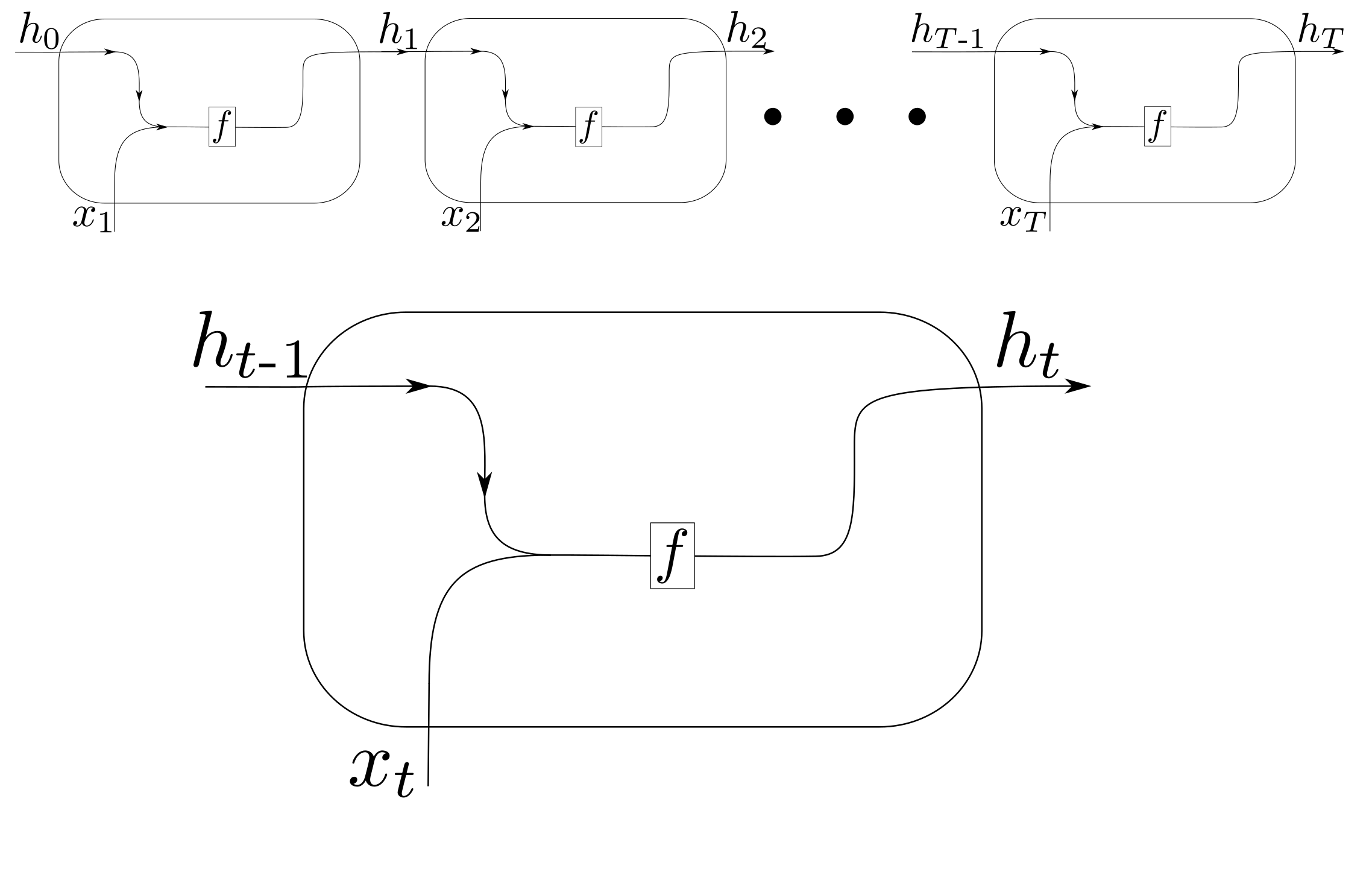}
 \caption{Top: The information flow of vanilla RNN. Bottom: A vanilla recurrent cell. Only the previous state and the input are needed to update the state.}
 \label{fig:vanilla_rnn}
\end{figure}

The cell operates over each element of the sequence until no more data is available. The parameters $\mathbf{W}$, $\mathbf{U}$ and $\boldsymbol{\mathit{b}}$ are shared across all time steps. They are learned during training using the Back Propagation Through Time (BPTT; \citealt{werbos1990backpropagation}) algorithm. 

In this form, the cell described in Equation~\ref{eq:vanila_rnn} and Figure~\ref{fig:vanilla_rnn} cannot treat long-term dependencies and presents numerical issues that limit its usability \citep{Bengio1994}, in part because the entire hidden state is being exposed to the new data. For long sequences, the information encoded in the hidden state at the start of the sequence might be lost at its end, because of the vanishing gradient problem which arises from the BPTT. At each iteration, the gradient must be computed and accumulated in each sequential step. The longer the sequence, the smaller the gradients can become, getting closer to zero at the start of the sequence. This nearly zero gradient means that the network cannot learn from all the data, thus not being appropriately trained.

Modifications have been applied to handle the vanishing gradient problem, Long Short Term Memory (LSTM; \citealt{Hochreiter1997}) and Gated Recurrent Unit (GRU; \citealt{Bahdanau2014}) cells being the dominant ones. Both cells implement multiplicative gates which alleviate the vanishing gradient problem and increase their long term memory. 

In our case of study, we choose GRU over LSTM, as it presents roughly the same performance \citep{Chung2014}, has fewer parameters and needs to save less information per time series. The LSTM cell needs two vectors to be stored, the cell state and hidden state, which are needed at each time step. To classify millions of objects every day, the model needs to request as little data as possible for every object. 

\begin{figure}
\centering
 \includegraphics[width=0.45\textwidth]{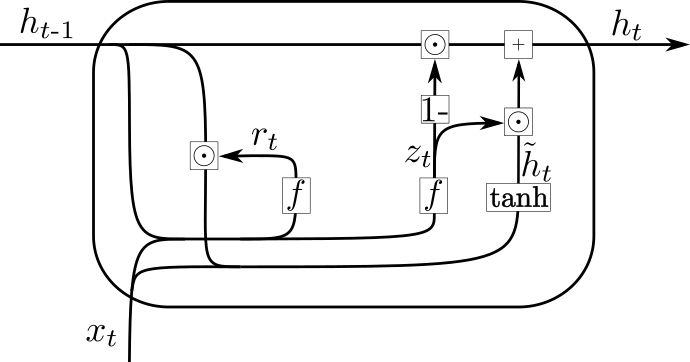}
 \caption{Gated Recurrent Unit cell diagram. The arrows represent the data flow. The labeled quantities are the same as in equations \eqref{eq:vanila_rnn}-\eqref{eq:hidden_satate}.}
 \label{fig:gru}
\end{figure}

The GRU cell information flow is shown in the diagram in Figure~\ref{fig:gru} and explained below. At the start of the step, the reset gate is computed as
  \begin{align}
     \boldsymbol{\mathit{r}}_\text{t} &= f \left ( \mathbf{W}_\text{r} \boldsymbol{\mathit{x}}_\text{t} + \mathbf{U}_\text{r} \boldsymbol{\mathit{h}}_\text{t-1} + \boldsymbol{\mathit{b}}_\text{r}\right ),  \label{eq:r_t}
 \end{align}
 which is used to expose part of the incoming hidden state $\boldsymbol{\mathit{h}}_\text{t-1}$, via a Hadamard product $\odot$, to compute a proposed hidden state
 \begin{align}
      \tilde{\boldsymbol{\mathit{h}}}_\text{t} &= \tanh{\left (\mathbf{W}_\text{h}\boldsymbol{\mathit{x}}_\text{t} + \mathbf{U}_\text{h}(\boldsymbol{\mathit{r}}_\text{t}\odot \boldsymbol{\mathit{h}}_\text{t-1}) + \boldsymbol{\mathit{b}}_\text{h}\right )}. \label{eq:hhat_t}
 \end{align}
 Then the update gate is computed as follows: 
 \begin{align}
 \boldsymbol{\mathit{z}}_\text{t} &= f \left ( \mathbf{W}_\text{z} \boldsymbol{\mathit{x}}_\text{t} + \mathbf{U}_\text{z} \boldsymbol{\mathit{h}}_\text{t-1} +\boldsymbol{\mathit{b}}_\text{z}\right ), \label{eq:z_t}
 \end{align}
which controls at what degree the proposed and the past hidden state will be combined, as an element-wise linear interpolation: 
\begin{align}
     \boldsymbol{\mathit{h}}_\text{t} &= (1-\boldsymbol{\mathit{z}}_\text{t})\odot \boldsymbol{\mathit{h}}_\text{t-1} + \boldsymbol{\mathit{z}}_\text{t} \odot\tilde{\boldsymbol{\mathit{h}}}_\text{t}.\label{eq:hidden_satate}
\end{align}

\begin{figure}
\centering
 \includegraphics[width=0.45\textwidth]{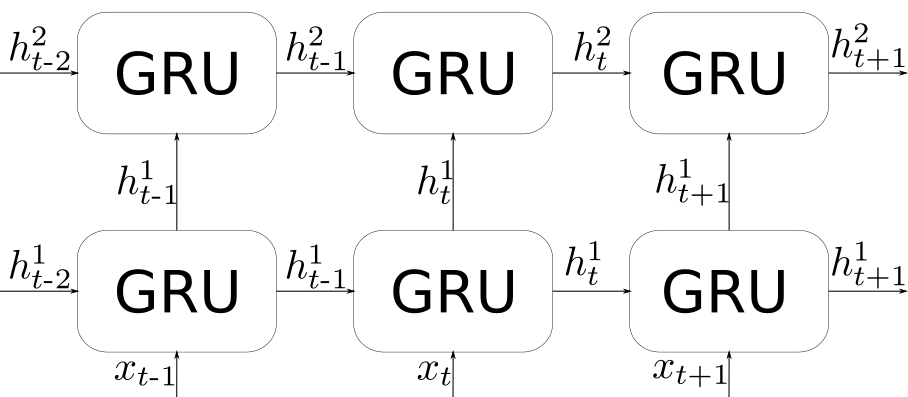}
 \caption{Stacked recurrent architecture. Each cell transmits its output hidden state to the next recurrent layer and to the next time setp.}
 \label{fig:multilayer}
\end{figure}

To increase representation complexity and generalization potential, additional recurrent layers can be stacked \citep{pascanu2014construct}. The second layer generates another sequence of hidden states, and instead of using the sequence elements as inputs, it uses the hidden states of the previous recurrent layer as inputs. This is shown in Figure~\ref{fig:multilayer}.

At the end of $T$ steps, the value of $\boldsymbol{\mathit{h}}_\text{T}$ will act as a representation of the entire time series. It is similar to a feature extraction method which takes an unstructured dataset and projects it into a fixed-length vector. 
This vector is the input of the next stage of the network, where it is transformed via fully-connected layers, which combine the information to produce a more informative representation. 

In a classification task, the result of the fully-connected layers must be projected to a vector whose dimension is the desired number of classes $K$. This projection needs to be transformed again, as it contains negative values and is not normalized.
The softmax function is used in this scenario, shown in Equation \ref{eq:softmax}. It takes a $K$-dimensional vector and transforms each of its $k$ components to values between \num{0} and \num{1}, which add up to 1: 

\begin{equation}
  \sigma(\boldsymbol{y})_k =
    \frac{\text{e}^{y_k}}{\sum_{j=1}^K{\text{e}^{y_j}}}.
 \label{eq:softmax}
\end{equation}

The output for an object can be interpreted as the probabilities of belonging to each class. Additionally, it is used to evaluate the cross-entropy, the classification loss function. This function takes the softmax output as input and compares it to the ground truth, which is expressed as a one-hot vector. This encoding creates a vector of length $K$ which sets to 1 the position corresponding to the class and 0 elsewhere. The cross-entropy can be written as:
\begin{equation}
 H(p,q) = -\sum_{k}{p(k)\log{q(k)}},
\end{equation}
where $p$ are the ground truth values, and $q$ the result of the softmax layer. This expression is differentiable, which makes it useful in a backpropagation training scheme. 
   
\section{Method}\label{section:Method_Description}
In this section, we explain the method used to preprocess the data and train our algorithm. Our main objective is to reduce the computational cost of the preprocessing stage, either for update a light curve with new observations or process a new light curve.

\subsection{Preprocessing}
Single-band light curves contain time as Modified Julian Date, magnitude, and the uncertainty associated with each measurement. Different objects have a different number of observations. Magnitudes depend not only on the intrinsic brightness of the object but also on the distance and the extinction. On top of that, different sources of noise such as weather, sky brightness, and crowding can render observations useless or prevent them at all. 

Comparing observations among different light curves is infeasible, as each object is measured at different times. A preprocessing step is done to make the information comparable. Feature-based methods usually preprocess each light curve by computing a series of descriptors and metrics derived from the time-series and their metadata. If new observations are obtained, some features usually must be recomputed to update them.

In our approach, we sort the light curve by time and compute the difference with previous measurements both in time and magnitude. It is worth to notice that generally the data is already sorted in time. In a real-time classification scheme, the data will arrive in temporal order so that no sorting will be required.

\begin{figure}
\centering
 \includegraphics[width=0.45\textwidth]{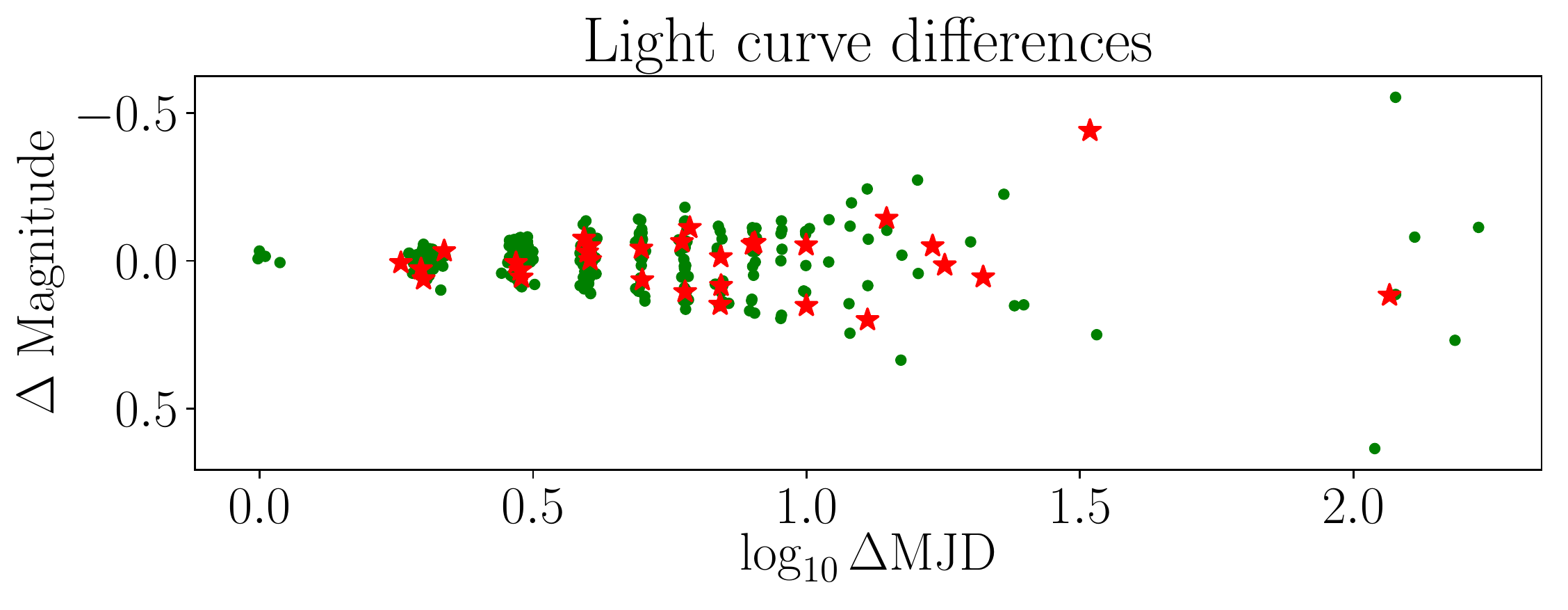}
 \caption{Preprocessing visualization for a Mira variable from OGLE-III. The $x$-axis shows the logarithm of the time differences, given in units of Modified Julian Days. Since Miras are periodic, the mean $\Delta \text{ Magnitude}$ is centered at \num{0}. The observed pattern in the time axis is a footprint of the survey cadence. The red stars correspond to observations of a single sliding window, which are spread all over the graph, exemplifying the random sampling of the light curves.}
 \label{fig:normalization}
\end{figure}

The differences utilize vector operations, requiring little computational overhead. For periodic objects, this process scales the magnitude around \num{0}. It is similar to subtracting the mean and helps the convergence of the network while maintaining the variability information, as shown in Figure \ref{fig:normalization}.
The process is illustrated in Equation~\ref{eq:preproc_0}. The first row is always eliminated since its difference cannot be computed, as no previous observations are available: 

\begin{align}
\begingroup
\setlength\arraycolsep{2pt}
\begin{bmatrix}
    t_1       & m_1\\
    t_2       & m_2\\
    t_3       & m_3\\
    \vdots & \vdots \\
    t_N       & m_N
\end{bmatrix}
-
\begin{bmatrix}
    0         & 0\\
    t_1       & m_1\\
    t_2       & m_2\\
    \vdots & \vdots \\
    t_{N-1}       & m_{N-1}
\end{bmatrix}
\rightarrow
\begin{bmatrix}
\hbox to 2pt{\vrule width 35pt depth -2.5pt height 3.0pt\hss}
    t_1         & m_1\\ 
    \Delta t_2       & \Delta m_2\\
    \Delta t_3       & \Delta m_3\\
    \vdots & \vdots \\
    \Delta t_N       & \Delta m_N
\end{bmatrix}
=
\begin{bmatrix}
    \Delta t_2       & \Delta m_2\\
    \Delta t_3       & \Delta m_3\\
    \vdots & \vdots \\
    \Delta t_N       & \Delta m_N
\end{bmatrix}. 
\endgroup
\label{eq:preproc_0}
\end{align}

We experiment using time and magnitude differences explicitly as inputs (scenario A), and using just the magnitude differences (scenario B), without explicitly using the time, but preserving the temporal order of the magnitude differences. 

Each row $i$ of the resulting matrix is given by the following equations for scenarios A or B, respectively:
\begin{align}
 {x}_{i}^A &= [\Delta t_i,\Delta m_i] \label{eq:tm_1},\\
 {x}_{i}^B &= [\Delta m_i] \label{eq:tm_2}.
\end{align}

In our experimental setup, including one observation at each step is detrimental since light curves can have hundreds of observations. It can hurt performance and convergence, as such long-term dependencies cannot be handled even for a GRU.

\begin{figure}
\centering
 \includegraphics[width=0.45\textwidth]{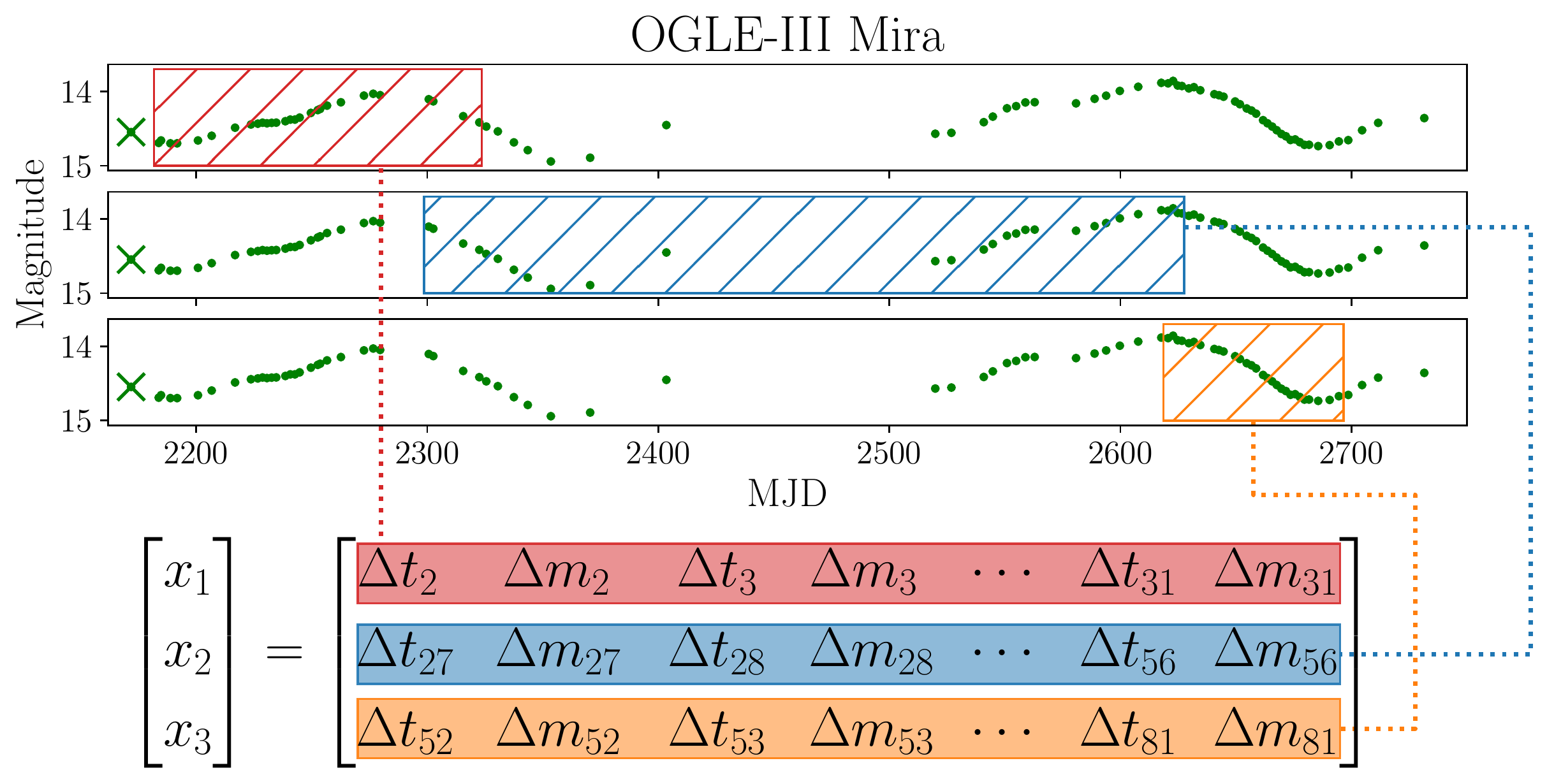}
 \caption{Sliding window sampling for a Mira variable from the OGLE-III dataset. In this visualization, the window size $w$ is \num{30} points and the stride $s$ is \num{25} points. Each row of the matrix represents the information obtained from each rectangle in the light curve, in sequential order. The time interval which each rectangle covers varies as a consequence of the non-uniform cadence. The first point, marked with a cross, is always dropped.}
 \label{fig:window}
\end{figure}

Instead, we employ a sliding window sampling scheme as done by multiple works \citep{Krizhevsky2012, Valenzuela2017} to handle long sequences, grouping observations, as seen in Figure \ref{fig:window}.

We concatenate the first $w$ rows of the matrix in Equation~\ref{eq:preproc_0} into a single vector. Then we skip $s$ rows and take another $w$ rows. The window $w$ and the stride $s$ are chosen to obtain a good classification and to control the number of rows in the matrix representation.

The number $M$ of vectors created following the procedure is expressed as
\begin{equation}
    M = \left\lfloor \frac{L - w + s}{s}\right \rfloor,
    \label{eq:M}
\end{equation}
where $L$ is the length of a light curve. We drop the last observations that cannot complete a row, as the network cannot process them. 

For every object, each of these vectors is stacked row-wise. The resulting matrix size is $M \times 2w$ for scenario A and $M \times w$ for scenario B.

Each row of the matrix representation, as shown in Figure \ref{fig:window}, will be used to feed one step in the recurrent portion of the network. 

This preprocessing helps to control the number of recurrent steps that need to be computed, limiting the effect of the vanishing gradient problem. Moreover, it enables us to train the network while repeating part of the previous information at each time step, which helps to improve the learning. 

The main difference with the preprocessing step of \citet{mahabal2017deep} is that we only compute the difference with the previous observation, and represent the light curve as a matrix. It enables our method to update the representations with a reduced computational cost. 

For the OGLE-III dataset, we employ a window size of \num{50} and a stride of \num{25}. For the other surveys, we use a window size of \num{4} and a stride of \num{2}. 

In all experiments, we discard objects with $M$ larger than \num{80}. We do not trim or sub-sample the light curves, as we cannot assume a priori which points in the light curve are more informative than others. The impact of removing them from the training sets is negligible since these light curves are few and are not representative of the data. 

\subsection{Model}
Our model is designed to be scalable to new observations of an object. In this way, the model needs $w$ new observations to update the hidden state and perform classification, not the entire light curve. For our purposes, the propagation of information is only forward instead of bidirectional \citep{schuster1997bidirectional}; otherwise, the network would need the entire light curve to update the hidden state. A unidirectional RNN can hurt performance but enable us to adapt our model to work in streaming scenarios, where new observations are added constantly, avoiding the re-computation as required in the case of traditional features.

We use two recurrent layers to increase complexity and overall generalization of the learned parameters. For each matrix representation, we feed one row at each cell, as described in Section~\ref{section:Background_Theory}.

We apply fully connected layers to increase the complexity of the representation extracted by the recurrent portion of the network and to project the final layer to the number of subclasses used. We apply dropout \citep{Srivastava2014} to these layers with a probability of \num{0.4} to avoid overfitting.

Finally, we apply a softmax layer. We consider each subclass as independent, letting the network learn without any imposed bias.
We use the Adam optimizer \citep{Kingma2014} a version of the Stochastic Gradient Descent (SGD) algorithm, which proves adequate for our objective.

\begin{figure}
\centering
 \includegraphics[width=0.5\textwidth]{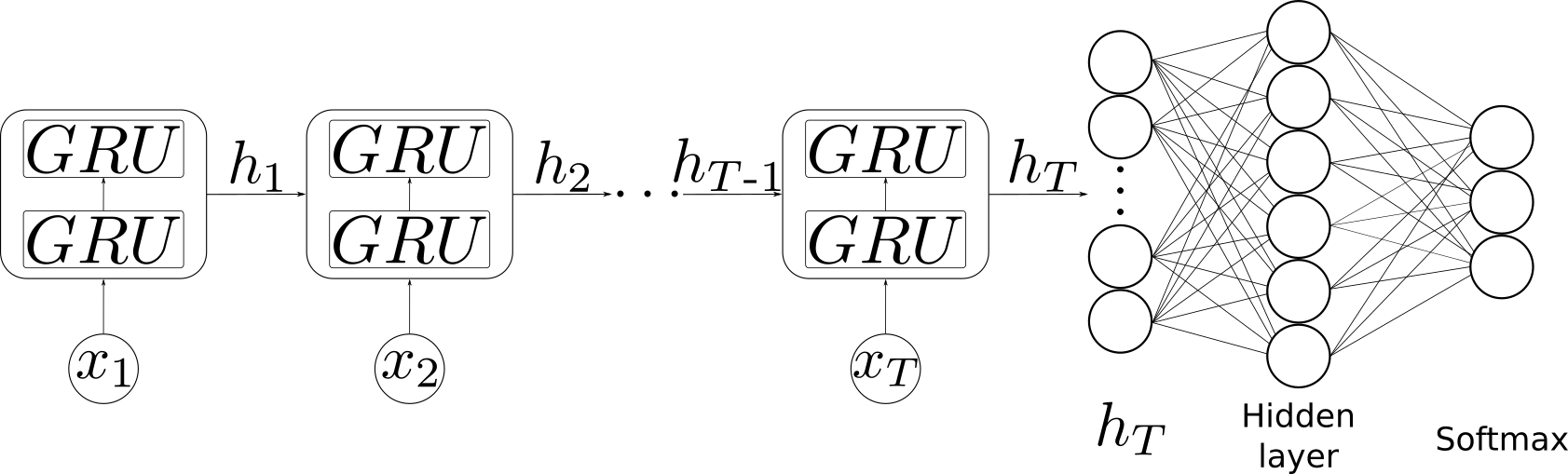}
 \caption{Proposed model. Each cell consist of two GRU cells. }
 \label{fig:model}
\end{figure}

A diagram of the proposed model is presented in Figure \ref{fig:model}. Our model differs from \cite{Naul2018recurrent}, as they use an encoder-decoder to extract features and an RF to perform classification. Contrary to our model, their encoder uses a bidirectional RNN. This means that, in order to extract features of a light curve with new observations, the model needs to analyze the entire light curve again. To explore massive catalogues which are continuously being updated, redoing the analysis can become infeasible in the long run. 

\subsection{Computational complexity}\label{section:Computational_Complexity} 
The computational complexity of all terms in a GRU cell, in equations (\ref{eq:r_t})-(\ref{eq:hidden_satate}), is
\begin{align}
O(\boldsymbol{\mathit{r}}_\text{t}) &= O(\boldsymbol{\mathit{z}}_\text{t})= O( \tilde{\boldsymbol{\mathit{h}}}_\text{t} )= O(2h^2 + 4hw+ 2h),& \\
O(\boldsymbol{\mathit{h}}_\text{t}) &= O(4h),& 
\end{align}
where $h=|\boldsymbol{\mathit{h}}_\text{t}|$. The total computational complexity of a single evaluation of the cell is
\begin{equation}
O(\text{GRU}) = O(6h^2 + 12hw+ 10h) \sim O(h^2 + 2hw).
\label{eq:GRU_complexity}
\end{equation}

For a light curve with a new set of observations, the computational cost remains the same as for one cell. It is proportional to the size of the hidden state squared and goes linearly with the size of the window. The cost of processing new observations of a light curve does not depend on its size: this is fixed by the parameters of the model.

For a new light curve, each cell will be evaluated $M$ times. In this scenario, the computational complexity is
\begin{equation}
O(\text{LC})  \sim O(M \times [h^2 + 2hw]) \sim O(L \times [h^2 + 2hw]).
\end{equation}
The value of $M$ is proportional to the number of observations $L$. This means that the complexity of a new light curve scales linearly with the number of observations it contains.

A bidirectional RNN cannot scale linearly as its unidirectional counterpart, as it requires to analyze the entire light curve, even if few observations are included \citep{graves2013speech}. Thus, the computational cost for a new set of observations and a new light curve is the same, being proportional to the total number of observations. It does not scale in a scenario where the light curves are being updated constantly, and it requires to process all the observations of an object.

\subsection{Training scheme}
Astronomical datasets keep growing over time, and so do the training sets themselves. Thus, training models in a single machine will not be possible. To prepare our model for larger training sets, we use techniques of distributed computing. We serialize the information of each object, which is comprised of its matrix representation and its class. This enables us to implement a variety of functions that help to feed the data as well as to manage the input pipeline efficiently.

At the beginning of the training phase, a fraction of the total number of objects is read and shuffled. From this portion, we extract a smaller part or batch. Batch training improves training speed, as with each batch, the network converges stochastically to the solution. Compared to the entire dataset, its size is manageable for the local hardware. Additionally, it helps to avoid local minima as each batch will be different due to shuffling. Its size depends on the specific problem and is usually set by experimentation. A small batch will be computed faster and make each example impact more the value of the parameters. 

Each time the entire training set has been passed through the network, we consider one epoch has passed. In batch training, the network will take multiple epochs to arrive at the solution, as the parameters slowly converge to their optimal values.

To feed the model, we pad each batch with rows filled with zeros, to match the number of rows of the largest matrix in that batch. For each object, we compute just the necessary $M$ recurrent steps for that object. Otherwise, the hidden states would be updated with unrelated information. This way, we can process objects with a different number of observations.

For each object, the model will predict the probability to belong to each class, as in Equation~\ref{eq:softmax}. We select the class with maximum probability to obtain a prediction. We measure the overall classification error $E$ as 
\begin{equation}
 E = 1 - \frac{\text{Correct predictions}}{\text{Total number of examples}}\label{eq:error}.
\end{equation}

\subsection{Experiments}\label{subsection:Experiments}
For each dataset, we divide our experimental setup into two parts. The first is designed to evaluate the performance of scenarios A and B. The second experiment aims to obtain the best classification using the architectures obtained in the firsts experiments. 

\subsubsection{Model selection}\label{subsubsection:experiments_architecture}

For each experiment in Table~\ref{tab:experiments_1}, we split each dataset into three parts stratified by the subclass. The training set is comprised of $70\%$ of the data, whereas $10\%$ of the latter is used to validate our results at every epoch. Finally, the remaining $20\%$ was used to test the error of our model. 

The training and validation sets are used to train and evaluate the convergence of the models. The test set is never involved in the training stage and is only used to assess the ability of the resulting model to generalize from the set of examples used to fit the model parameters (the training set). The stratification is needed since the subclasses are not represented equally, due to the intrinsic timescales, and the survey's observational constraints. Thus, all the splits must maintain the same proportion of labels. Even with stratified labels, some classes are still orders-of-magnitude more numerous than others. 

Models trained with all the available data will be biased towards the most numerous classes, in detriment of the less-represented ones. To alleviate this issue, in each experiment, we sample randomly without replacement up to \num{40000} elements from each category. This number is large enough to get sufficient diversity from each class, and not too large to bias the classifier. The sampling is done in each training instance. This process also reduces the computational cost of the training without a loss in classification performance.

To select the optimal model, we explore four different architectures which will be trained and evaluated to obtain an optimal model. These are presented in Table~\ref{tab:experiments_1}.

In each experiment, we use four different dimensions of the hidden state $h = (25,50,100,200)$. 
To standardize the experiments, we set the size of the fully connected layers to the double of the hidden state size. In total, we evaluate 16 different architectures. We run each experiment 10 times, and average the results to obtain a representative score.

\begin{table}
\caption{We test four different architectures. FC Layer means the inclusion of an additional fully connected layer, for a total of two. "Time" implies the explicit use of the time information as shown in Equation~\ref{eq:tm_1}. }
\begin{center}
        \begin{tabular}{lccc} 
        \toprule
              & Scenario & Time & FC Layer  \\
        \midrule
         Exp 1 &\multirow{2}{*}{A}&  Yes & Yes  \\
          Exp 2 &  &Yes & No   \\
         Exp 3 &\multirow{2}{*}{B}&  No  & No   \\
         Exp 4& &  No  & Yes \\
        \bottomrule
        \end{tabular}
\end{center}
\label{tab:experiments_1}
\end{table}

Architectures in experiments 1 and 2 will have a bigger input vector than in experiments 3 and 4, which will impact the overall number of parameters to be learned as well as the inclusion of an extra hidden layer.

For each dataset, we choose the architecture and hidden state size with the smaller error in the test set, as a metric of the performance. If two networks achieve similar results, we choose the one with smaller input size. A smaller input means reduced computational complexity, as well as requiring less information to update the hidden state, which is essential for a scalable model.

\subsubsection{Best model evaluation}

To evaluate the best models in the second set of experiments, we perform 5-fold cross-validation. First, as a validation set, we separate a stratified sample containing $10\%$ of the data, to validate each fold equally. Then, we split the remaining $90\%$ of the data into five stratified folds, each one consisting of $80\%$ for training, and $20\%$ of the remaining data for testing. 
As with the previous set of experiments, we sample \num{40000} objects per subclass.

The best model found for each dataset is trained with a lower learning rate and \num{2000} epochs. We stop the training when the network reaches \num{2000} epochs or starts to overfit, which occurs when the validation accuracy starts to increase.

\section{Datasets}\label{section:Datasets}
Many catalogues of variable stars are available in the literature, but few have sufficient labels and labelled samples to train our model. Here we detail the ones used in this work.

\subsection{Optical Gravitational Lensing Experiment (OGLE)}
One of the most extensive datasets available is provided by the OGLE survey \citep{Udalski2004}, which in its third phase operated from \num{2001} to \num{2009}. It observed the Galactic bulge, disk, and both Magellanic Clouds, and classified more than 450 thousand variable stars. The observations were conducted in the $V$ and $I$ bands, with the latter having ten times more observations than the former. This catalogue represents a clean sample as experts confirmed all objects. Furthermore, the light curves are well defined with high signal-to-noise ratios in the majority of the observations.
 
The OGLE-III catalogue does not include non-variable (NonVar) stars. To obtain a sample, we use the available photometry of OGLE-II, limiting the maximum amplitude to \num{0.015}~mag. We extract at most \num{20000} such stars per observed field, or \num{52619} in total, before imposing any constraints over the data. 

In the training phase, we impose a minimum of \num{500} examples per category. Some classes contain well-represented subclasses, such as RRab and RRc in the case of RR Lyrae stars, while some classes do not. The latter, which includes classical Cepheids, were labelled as single classes, without for instance differentiating between fundamental and first overtone pulsators.
 
The categories and their corresponding numbers of stars for the OGLE-III catalogue are displayed in Table~\ref{table:OGLE-numbers}. In this table, the eclipsing variables (ECL) are subdivided into contact (EC), semi-detached (ESD), and detached binaries (ED). The Long Period Variables (LPV) include Miras, Semi Regular Variables (SRVs), and OGLE Small Amplitude Red Giants (OSARGs). RR Lyraes (RR Lyr) are subdivided into RRab and RRc subclasses. Type I Cepheids (Cep) and $\delta$~Scuti (DSct) variables are also included.

\begin{table}
\centering
\caption{Total numbers of elements per class in the OGLE-III dataset.}
  \begin{tabular}{@{}cS}
    \toprule
    \textbf{Class} & \textbf{Number} \\
    \midrule
      EC & 6862\\
      ESD & 9475 \\
      ED & 21503 \\
      OSARG & 234932 \\
      SRV & 34835 \\
      Mira & 6090 \\
      RRab & 25943\\
      RRc & 7990\\
      Cep & 7836 \\
      DSct & 2822 \\
      NonVar & 34815\\
    \bottomrule
      Total & 393103 \\
  \end{tabular}
  \label{table:OGLE-numbers}
\end{table}

\subsection{Gaia DR2 Catalog of variable stars}

The Gaia mission \citep{prusti2016gaia,brown2018gaia} was designed to measure with detail positions, parallaxes (hence distances), proper motions, and physical characteristics of stars. Its light collecting area is \SI{0.7}{m^2}, and the collected photometry includes three bands, namely $G$, $G_{BP}$ and $G_{RP}$, reaching magnitude \num{21} in the $G$ band. Its data release 2 (DR2) delivered, among others, a catalog of \num{363969} variable stars (Eyer et al. 2019).

The light curves present in this dataset typically have low intrinsic noise and high signal-to-noise, biased towards bright objects. It still has a sufficient number of categories and number of examples to be used as a training set, although not representative of the real population of stars.
The published dataset was constructed using a multi-stage classifier, based on light curve features as well as parallaxes, metallicities, astrophysical relations, among other metadata. Thus, the classification based on light curves will not achieve a perfect score. 

We select stars with at least \num{20} observations and a minimum number of objects per class of \num{500}. The classes and number of stars for the Gaia data are displayed in Table~\ref{table:GAIA-numbers}. SX Phoenicis and $\delta$~Scuti stars are bundled into a single category (DSct\_SXPhe). RR Lyrae stars include subdivisions into RRab, RRc, and RRd subtypes. Also, two subclasses of Cepheids (Cep) are considered, Type I (T1Cep) and type II (T2Cep). It must be noted that in this dataset, RV Tauris are considered part of type II Cepheids.

\begin{table}
\centering
\caption{Total numbers of elements per class in the Gaia dataset.}
  \begin{tabular}{@{}cS}
    \toprule
    \textbf{Class} &\textbf{Number} \\
    \midrule
        T1Cep & 6274 \\
        T2Cep & 1308 \\
        DSct\_SXPhe & 5178 \\
        MIRA\_SR & 87818 \\
        RRab &  78049 \\
        RRc & 21116 \\
        RRd & 566 \\
    \bottomrule
        Total & 200669 \\    
  \end{tabular}
  \label{table:GAIA-numbers}
\end{table}

\subsection{Wide-field Infrared Survey Explorer (WISE)}
The Wide-field Infrared Survey Explorer (WISE, \citealt{Wright2010}) is a space telescope launched in 2009. Its purpose was to map the infrared sky at \SIlist{3.4;4.6;12;22}{\micro m} wavelength bands with a \SI{40}{\centi m} telescope. As the mission depleted its coolant, the last two bands were unable to perform science observations.

Especially for pulsating stars, infrared light curves are known to have smaller amplitudes than in the visible \citep[see, e.g.,][]{Catelan2015}. Thus, this dataset enables us to replicate the scenario of a new telescope trying to detect objects near the limit magnitude, where variable objects cannot easily (if at all) be identified by experts directly. Moreover, we want to leverage the benefits of machine learning to extract signal from noisy data.

To obtain the labels for the light curves, we perform a cross-match between OGLE-III LPVs \citep{soszynski2009optical, soszynski2011optical, soszynski2013optical}, OGLE-IV eclipsing binaries \citep{pawlak2016ogle, soszynski2016ogle_ecl} and RR Lyraes \citep{soszynski2016ogle_rrl, soszynski2014over}, and the Gaia catalogue of variable stars using the already available cross-match with WISE objects \citep{marrese2019gaia}. In addition to the previous classes described for OGLE-III and Gaia, NC corresponds to non-contact binaries from OGLE-IV.

For the OGLE data, we perform the cross-match using a matching radius of $\ang{;;1}$. We remove any object with two or more possible matches to privilege the purity of the sample. Furthermore, we remove objects present in both surveys but with different classifications. 

We select stars with at least \num{20} observations and set the minimum number of examples per category to \num{500}. WISE non-variable (NonVar) stars are obtained from the IRSA infrared science service. We query the Large Magellanic Cloud and the Galactic bulge. We impose a maximum value for the variability flag for the non-variable objects, to be \num{5} in the \textit{W1} band. This flag value is defined as most likely not variable. 
Additionally, we remove objects that were cross matched to the various catalogues of variable stars described at the beginning of the subsection but were categorized as non-variable according to the WISE variability flag.

The classes and number of stars for the WISE data are displayed in Table \ref{table:WISE-numbers}. 

\begin{table}
\centering
\caption{Total numbers of elements per class in the WISE dataset.}
  \begin{tabular}{@{}cS}
    \toprule
    \textbf{Class} &\textbf{Number} \\
    \midrule
        NC & 2237 \\
        Cep & 1884 \\
        OSARG & 53890 \\
        SRV & 8605 \\
        Mira &  1396 \\
        RRab &  16412 \\
        RRc & 3831 \\
        DSct\_SXPhe & 1098 \\
        NonVar & 32795 \\
    \bottomrule
        Total & 122148\\
  \end{tabular}
  \label{table:WISE-numbers}
\end{table}

\begin{figure}
 \includegraphics[width=0.45\textwidth]{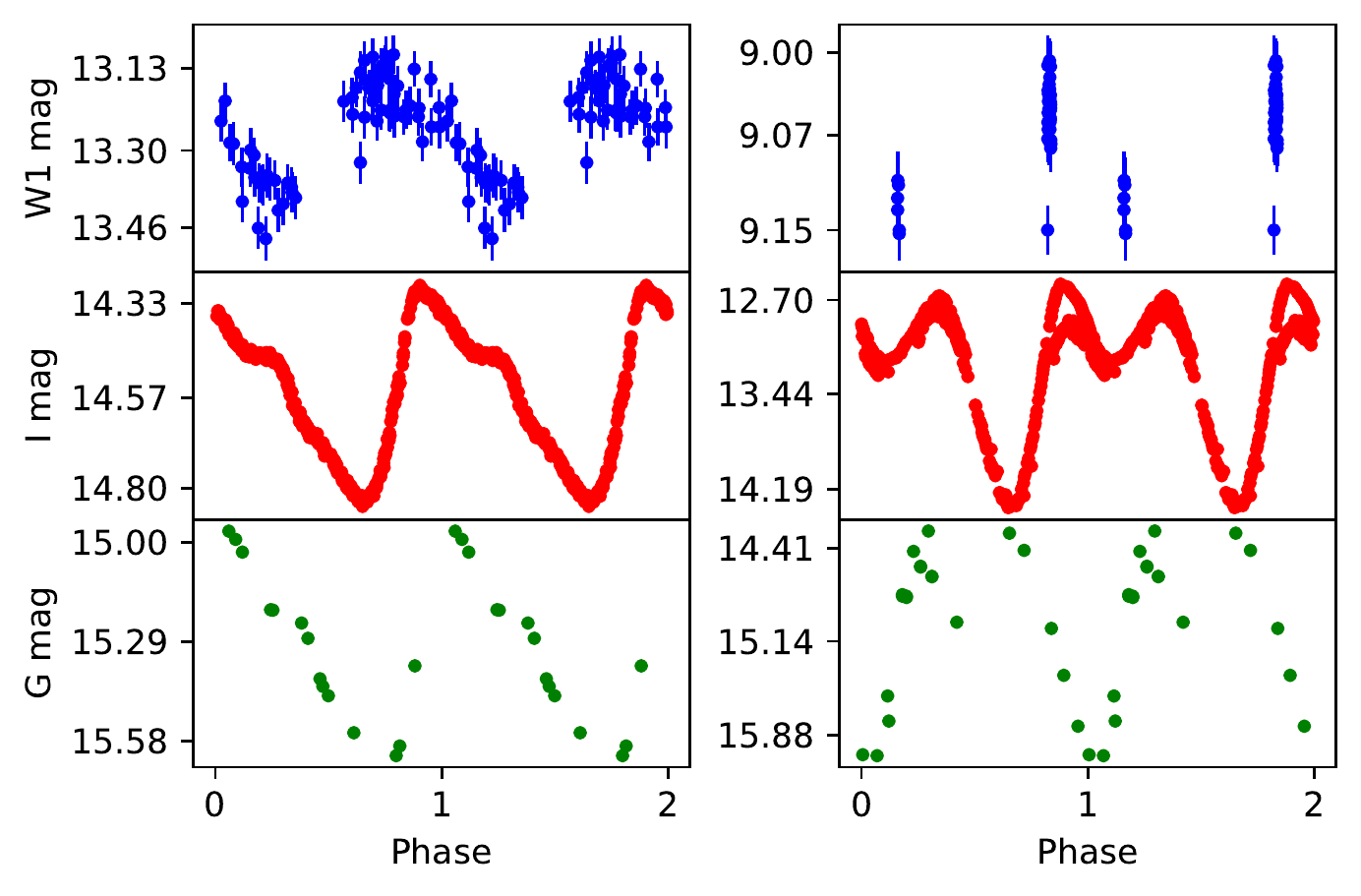}
 \caption{Folded light curves for the same objects, obtained by cross-match. From top to bottom, WISE, OGLE-III and Gaia. The sampling for each survey differs, as well as the observational uncertainty. We assume arbitrary zero epoch for each survey.
 Left: a type I Cepheid with a period of \num{7.23}~days. Right: a Mira with a period of \num{543.4}~days.
}
 \label{fig:Both_LightCurves}
\end{figure}

Figure~\ref{fig:Both_LightCurves} shows phased light curves for a Cepheid and a Mira that are present in all three surveys. The variability is seen in all light curves, although the noise characteristics, observational errors and cadence are different. Due to the observational pattern of WISE, objects with long periods such as Miras are frequently not observed throughout the entire pulsation cycle, which can result in poorly-sampled light curves. This behaviour can be expected in the early stages of new surveys. Since this is an aspect of the problem that we wish to address with our method, it is necessary to include this kind of incomplete time-series in our dataset. Naturally, such light curves present a challenge for any classifier.

\section{Results}\label{section:Results}
Our experiments aim to evaluate the classification accuracy in three different datasets, as described in Section~\ref{subsection:Experiments}. We test our models and compare them with RF and FATS features. 

We perform our experiments on a computer equipped with 16 GB of RAM, four cores and eight threads and a single GTX 1080 Ti graphics card. Our algorithm is developed using the TensorFlow library \citep{Abadi2016}.

We use the Scikit-learn library \citep{Pedregosa2011} for the implementation of the RF. The RF is trained with all the objects in each catalogue using the default parameters. The 59 single-band features are computed using the available Python 2 FATS package, which is implemented for CPU only. Many features could be parallelized in GPU, but at the time of writing, those were not available as a unified package.

\subsection{Model selection}
We now show the convergence of each experiment described in Section~\ref{subsubsection:experiments_architecture}. Each of the four experiments tested four different hidden state sizes, ten times. For the sake of clarity, we show the one corresponding to the best score in each experiment in Table \ref{tab:experiments_1}. Since we test the behaviour of each model, we are not optimizing other hyper-parameters, such as $w$ and $s$, dropout probability, learning rate, the parameters of the optimizer or the size of the fully connected layers.

\begin{figure}
 \includegraphics[width=0.5\textwidth]{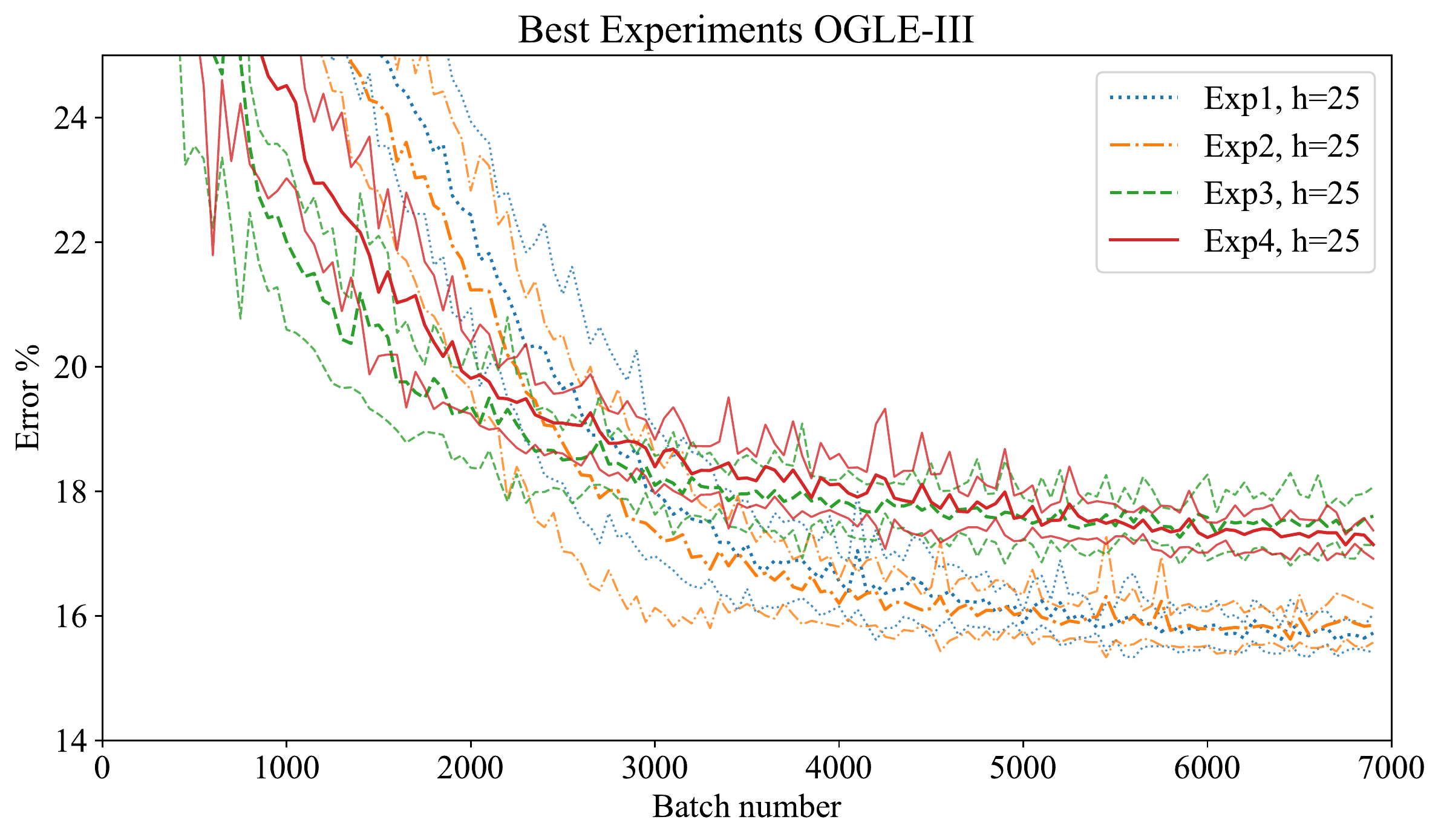}
 \caption{Curve of percentage error as a function of the number of epochs for each one of the experiments for the OGLE-III dataset. In all the experiments, the smaller embedding size performs better. The best result is obtained for Exp 1 at batch number \num{6700}. For each experiment, we plot the mean, plus and minus one standard deviation as a lighter shade.}  
 \label{fig:ogle_best_4exp}
\end{figure}

\begin{figure}
 \includegraphics[width=0.5\textwidth]{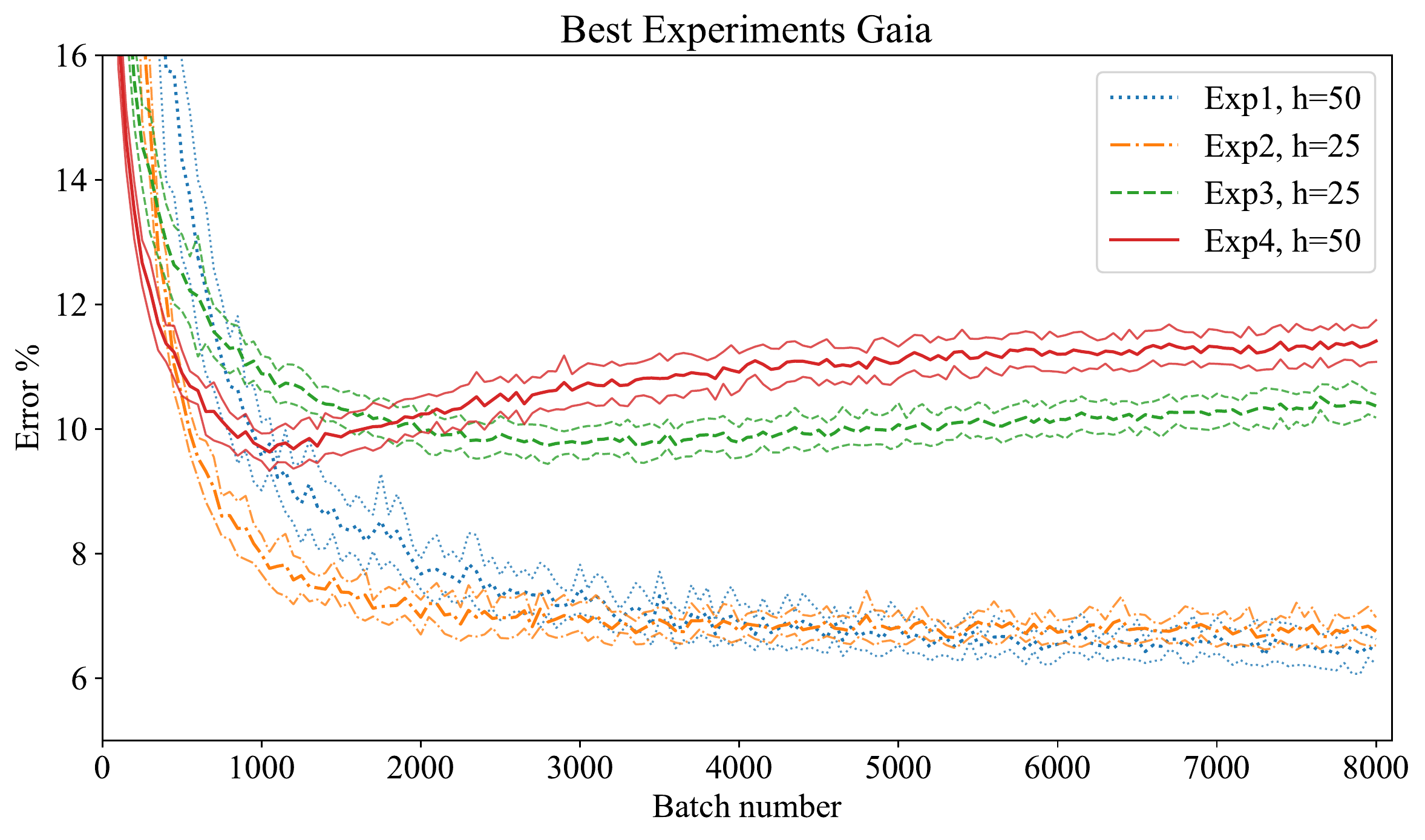}
 \caption{Curve of error as a function of the number of epochs for the best realization of experiments for the Gaia dataset. The best result is obtained for Exp 2 at batch number \num{7300} for a hidden state size of \num{50}. For each experiment, we plot the mean, plus and minus one standard deviation as a lighter shade.}
 \label{fig:GAIA_best_4exp}
\end{figure}

\begin{figure}
 \includegraphics[width=0.5\textwidth]{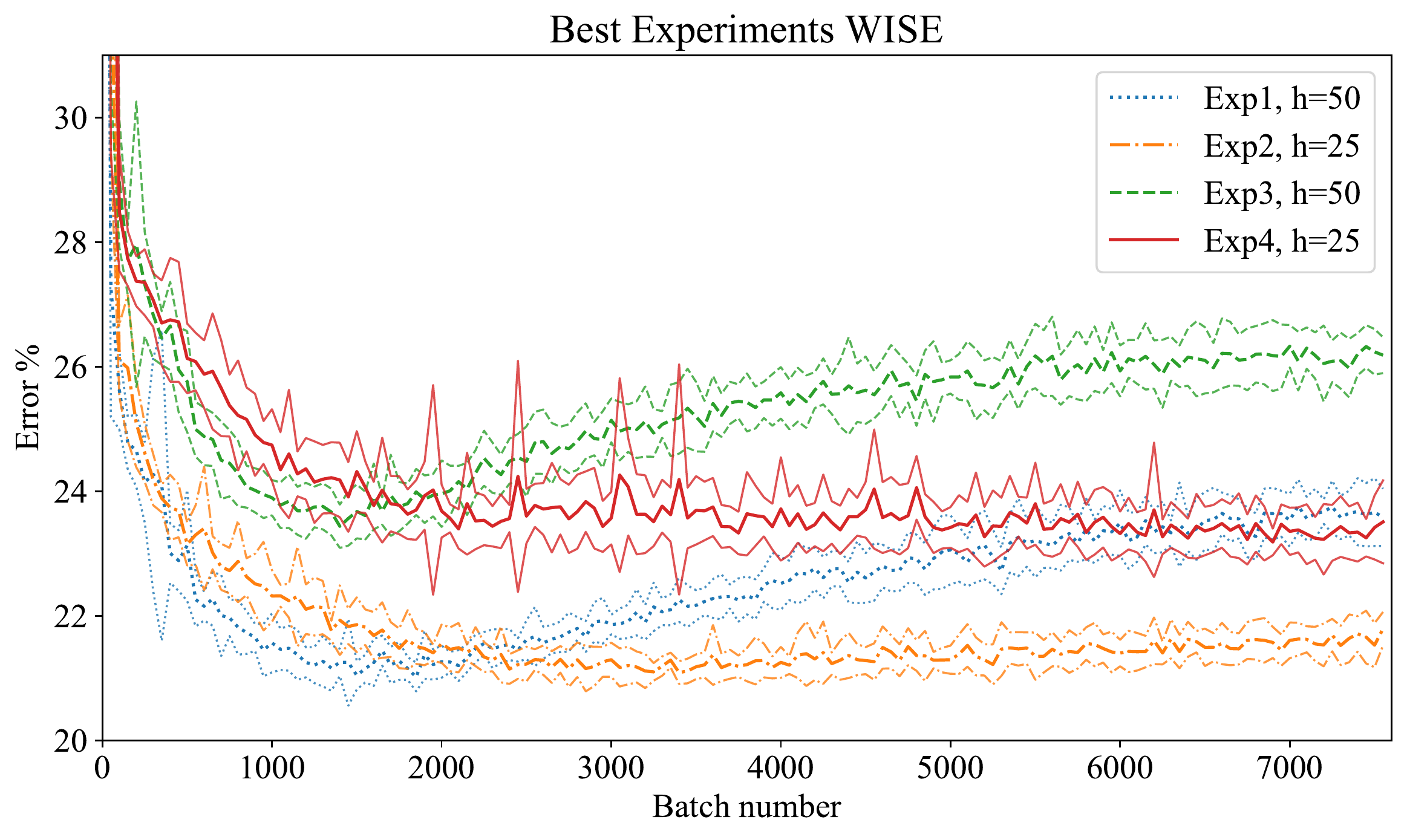}
 \caption{Curve of error as a function of the number of epochs for the best realization of experiments for the WISE dataset. The best result is obtained for Exp 1 at batch number \num{1850} for a hidden state size of \num{50}. For each experiment, we plot the mean, plus and minus one standard deviation as a lighter shade.}
 \label{fig:WISE_best_4exp}
\end{figure}

The initialization of the parameters plays an essential role in the training of neural networks \citep{glorot2010understanding}. At first, the parameters adjust quickly towards the optimal values, which can be seen in the first 50 epochs of Figures \ref{fig:ogle_best_4exp}, \ref{fig:GAIA_best_4exp}, and \ref{fig:WISE_best_4exp}, where the error diminishes fast. After that, the values slowly converge with
the subsequent parameter updates.


For OGLE-III, Figure~\ref{fig:ogle_best_4exp} shows that in all the experiments, models with a hidden state size of \num{25} perform better than models with larger representation sizes.
The models that include time explicitly perform $\sim1.5\pm 0.4$ per cent better than their counterparts. The inclusion of an additional hidden layer improves results marginally. The architecture which will be fine-tuned is the one corresponding to experiment 1.

For Gaia, Figure~\ref{fig:GAIA_best_4exp} shows that architectures that include time information explicitly perform $\sim3.5\pm0.3$ per cent better. Experiment 1 shows the best results for a hidden state size of \num{50}, marginally better than experiment 2.

For WISE, Figure~\ref{fig:WISE_best_4exp} also shows that including the time improves the results. Experiment 1 shows overfitting but obtains marginal better score than experiment 2. We opt to fine-tune the one corresponding to Experiment 1 with a hidden state size of \num{50}.

Although models tested in scenario B perform worse than the corresponding ones for scenario A, the performance degradation is relatively small, of order $~2$ per cent in all surveys. We can observe this effect in all tested surveys. Interestingly, the network can classify the time series without explicit time information. This indicates that the network is somehow able to successfully extract and process the limited supplied information, following a procedure that we have not yet been able to fully understand. The representation is robust enough to obtain an acceptable classification score, independent of the cadence of the survey.

In our tested models, the optimal hidden state sizes are no larger than \num{50}. We choose to fine-tune the best performer, even when the differences with other experiments are small.
We do not perform an exhaustive search for the optimal embedding size nor optimize the number of features for the RF. As such, we compare the representation size of both models equally, as they are of the same order of magnitude.

Our results show that the representation size does not need to be large in order to perform adequate classification. A small representation improves the generalization potential while also reducing inference and training times, as the required parameters to be learned are fewer. As such, our procedure can be deployed in consumer hardware while maintaining its ability to scale to bigger datasets, by increasing the number of GPUs instead of CPUs.

\subsection{Classification results}

\begin{figure}
 \includegraphics[width=0.5\textwidth]{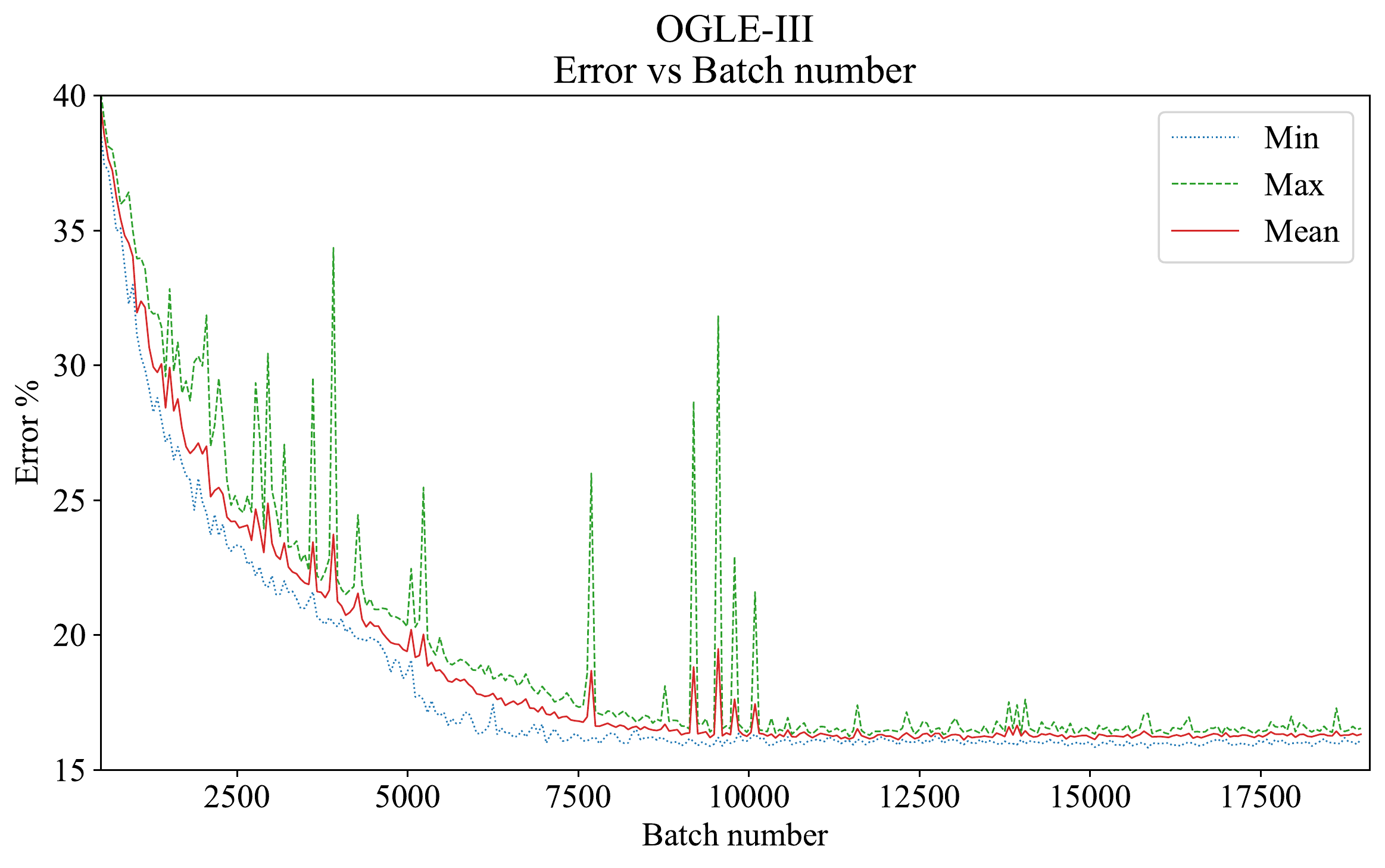}
 \caption{Curve of error as a function of the batch number for the best architecture for the OGLE-III dataset.  Further training does not improve the results, as the network gets stuck improving the classification of some subclasses in detrimento of the others.}
 \label{fig:OGLE_model}
\end{figure}

\begin{figure}
 \includegraphics[width=0.5\textwidth]{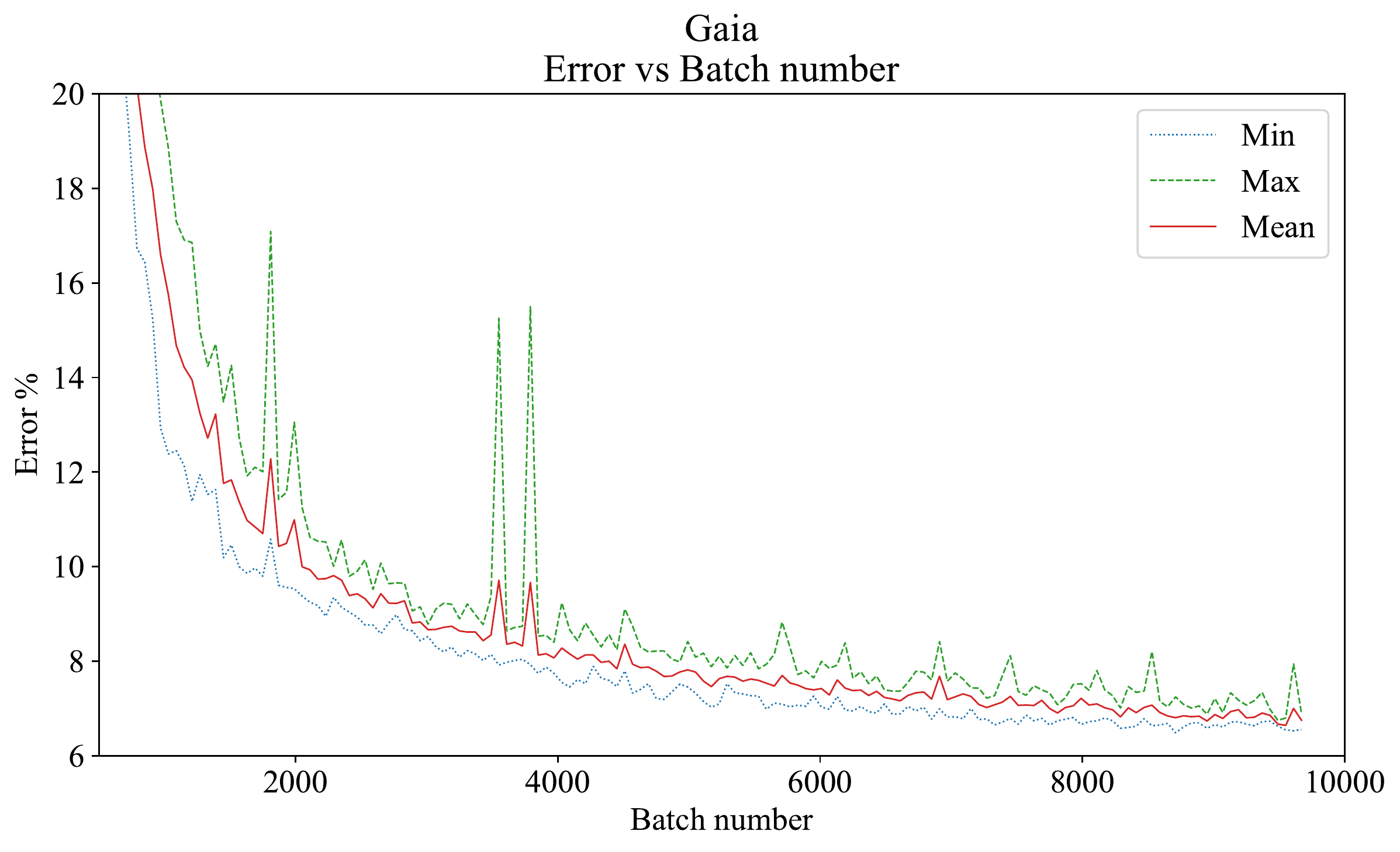}
 \caption{Curve of error as a function of the batch number for the best architecture for the Gaia dataset. The networks can be improved further, but the improvement would be smaller than a few percent.}
 \label{fig:GAIA_model}
\end{figure}

\begin{figure}
 \includegraphics[width=0.5\textwidth]{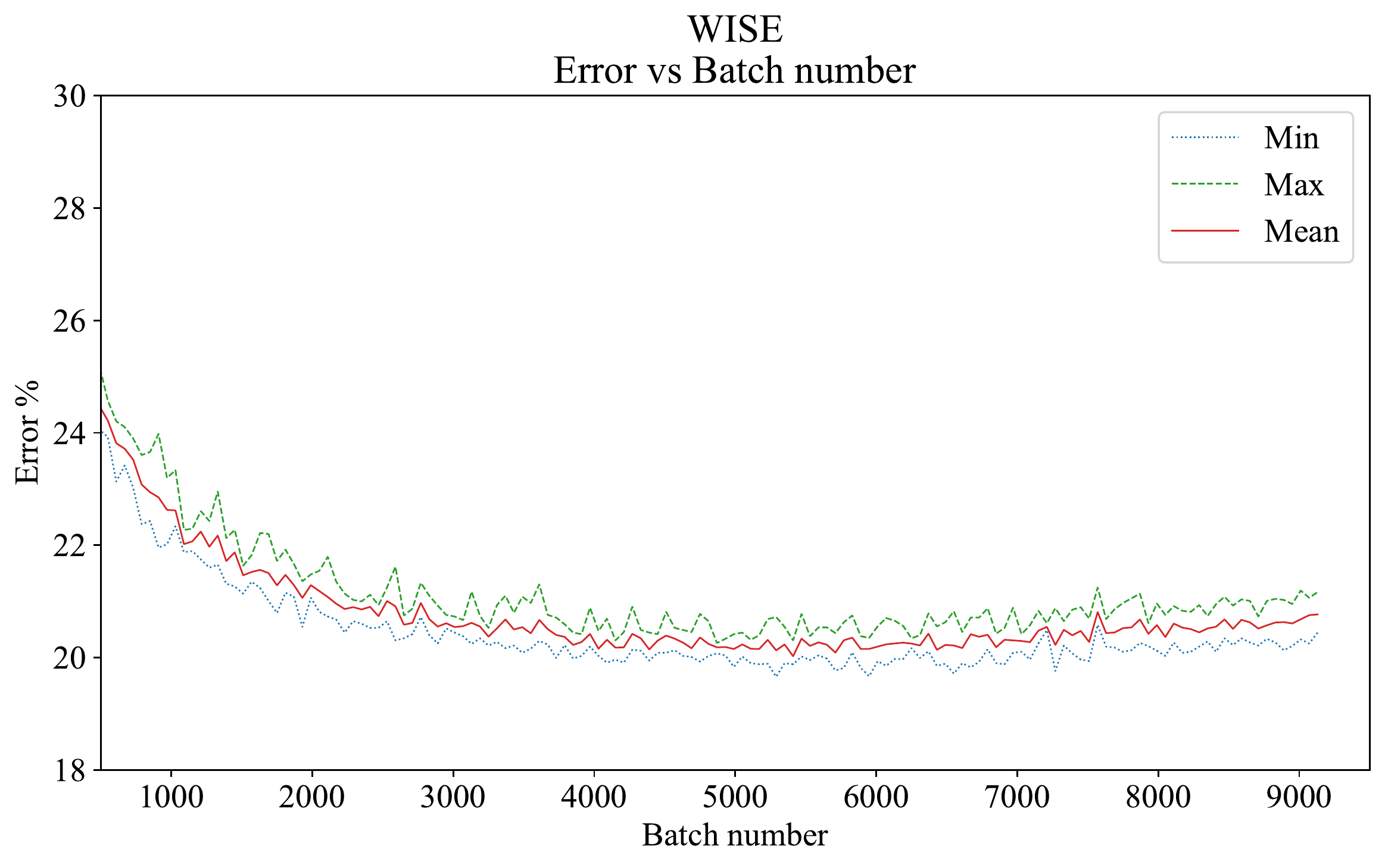}
 \caption{Curve of error as a function of the batch number for the best architecture for the WISE dataset. The model reaches a minimum around batch \num{5500}, then the validation error starts to increase. We consider our best the model up to the minimum.}
 \label{fig:WISE_model}
\end{figure}

Here we present the results of the cross-validation from the best models selected in the previous experiments. Figures \ref{fig:OGLE_model}, \ref{fig:GAIA_model}, and \ref{fig:WISE_model} show the evolution of the validation error as a function of the training batch number. We use batch number instead of epochs since we are looking for the best model, which can be obtained halfway of one training epoch. Each graph shows the minimum, maximum, and mean classification error of all folds, evaluated in their respective test set. The maximum batch number correspond to the \num{2000} training epochs. The maximum batch number will change depending on the batch size and the dataset size. The spikes are a consequence of momentum optimizers such as ADAM.

We compare our results with the RF classifier using \num{1000} trees. We perform the same 5-fold stratified cross-validation as well. We opt to use cross-validation instead of an out-of-the-bag estimate to compare it directly to our model.

The validation error in OGLE and Gaia decreases in the corresponding \num{2000} epochs. Further training would only improve the results marginally. For WISE, the situation is different, as the validation reaches a minimum at batch number \num{5390}. After this point, the model starts to overfit data, decreasing its generalization potential, as seen in Figure~\ref{fig:WISE_model}. For WISE, we finish the training at \num{2000} epochs, but consider only the model which obtains the minimum error in the validation set.

To get a class level score, we group the labels at the subclass level, as shown in Section \ref{section:Datasets}. For example, T1Cep and T2Cep are considered of the same Cep class. Then we compute accuracy and F-Score metrics. 

\begin{table}
    \centering
    \caption{Classification accuracy for the OGLE-III dataset.}
    \label{tab:classification_ogle}
    \begin{tabular}{lcc} 
        \hline
        Class & RNN & RF \\
        & Acc \& F-Score & Acc \& F-Score \\
        \hline
        Cep  & 0.72 - 0.69 & 0.97 - 0.97\\
        RR Lyr &0.90 - 0.91& 0.99 - 0.99\\
        Dsct&0.72 - 0.72& 0.93 - 0.95\\
        ECL & 0.94 - 0.94& 0.97 - 0.98\\
        LPV & 0.99 - 0.99 &1.00 - 1.00\\
        NonVar &1.00 - 1.00 & 1.00 - 1.00\\
        \hline
        Subclass & RNN & RF \\
        \hline
        Cep  & 0.72 - 0.69 & 0.97 - 0.97\\
        RRab & 0.85 - 0.80 & 0.99 - 0.99\\
        RRc & 0.30 - 0.40&  0.98 - 0.97\\
        Dsct& 0.72 - 0.72& 0.93 - 0.95\\
        EC  & 0.54 - 0.64 & 0.79 - 0.84\\
        ED  & 0.93 - 0.79 & 0.92 - 0.89\\
        ESD & 0.24 - 0.35 & 0.61 - 0.65\\
        Mira & 0.92 - 0.91 & 0.97 - 0.97\\
        SRV & 0.93 - 0.91 & 0.82 - 0.81\\
        OSARG & 0.90 - 0.92& 0.97 - 0.97\\
        NonVar &1.00 - 1.00 & 1.00 - 1.00\\
        \hline
    \end{tabular}
\end{table}

\begin{table}
    \centering
    \caption{Classification accuracy for the Gaia dataset.}
    \label{tab:classification_GAIA}
    \begin{tabular}{lcc} 
        \hline
        Class & RNN & RF \\
        & Acc \& F-Score & Acc \& F-Score \\
        \hline
        Cep & 0.78 - 0.83& 0.76 - 0.84\\
        Mira\_SR & 0.99 - 0.98& 1.00 - 0.99\\
        DSct\_SXPhe & 0.75 - 0.77 & 0.81 - 0.83\\
        RR Lyr &0.98 - 0.9&0.99 - 0.99\\
        \hline
        Subclass & RNN & RF \\
        \hline
        T1Cep & 0.81 - 0.82& 0.82 - 0.85\\
        T2Cep & 0.75 - 0.24& 0.19 - 0.30\\
        Mira\_SR & 0.99 - 0.98& 1.00 - 0.99\\
        DSct\_SXPhe & 0.75 - 0.77& 0.81 - 0.83\\
        RRab &0.96 - 0.96& 0.98 - 0.98\\
        RRc & 0.94 - 0.92& 0.95 - 0.94\\
        RRd &  0.00 - 0.00& 0.06 - 0.10\\
        \hline
    \end{tabular}
\end{table}

\begin{table}
    \centering
    \caption{Classification accuracy for the WISE dataset.}
    \label{tab:classification_wise}
    \begin{tabular}{lcc} 
        \hline
        Class & RNN & RF \\
        & Acc \& F-Score & Acc \& F-Score \\
        \hline
        ECL & 0.39 - 0.44&0.09 - 0.16\\
        RR Lyr & 0.96 - 0.93& 0.91 - 0.90\\
        DSct\_SXPhe & 0.03 - 0.06& 0.01 - 0.02\\
        Cep & 0.34 - 0.41& 0.06 - 0.10\\
        LPV & 0.96 - 0.94& 0.96 - 0.93\\
        NonVar &  0.89 - 0.90& 0.84 - 0.84\\
        \hline
        Subclass & RNN & RF \\
        \hline
        NC & 0.39 - 0.44&0.09 - 0.16\\
        RRab & 0.95 - 0.84& 0.93 - 0.83\\
        RRc & 0.07 - 0.12& 0.05 - 0.10\\
        DSct\_SXPhe & 0.03 - 0.06& 0.01 - 0.02\\
        Cep & 0.34 - 0.41& 0.06 - 0.10\\
        SRV & 0.45 - 0.49& 0.56 - 0.60\\
        Mira &  0.04 - 0.07& 0.43 - 0.54\\
        OSARG &  0.89 - 0.84& 0.92 - 0.87\\
        NonVar & 0.89 - 0.90& 0.84 - 0.84\\
        \hline
    \end{tabular}
\end{table}

\begin{figure*}
 \includegraphics[width=\textwidth]{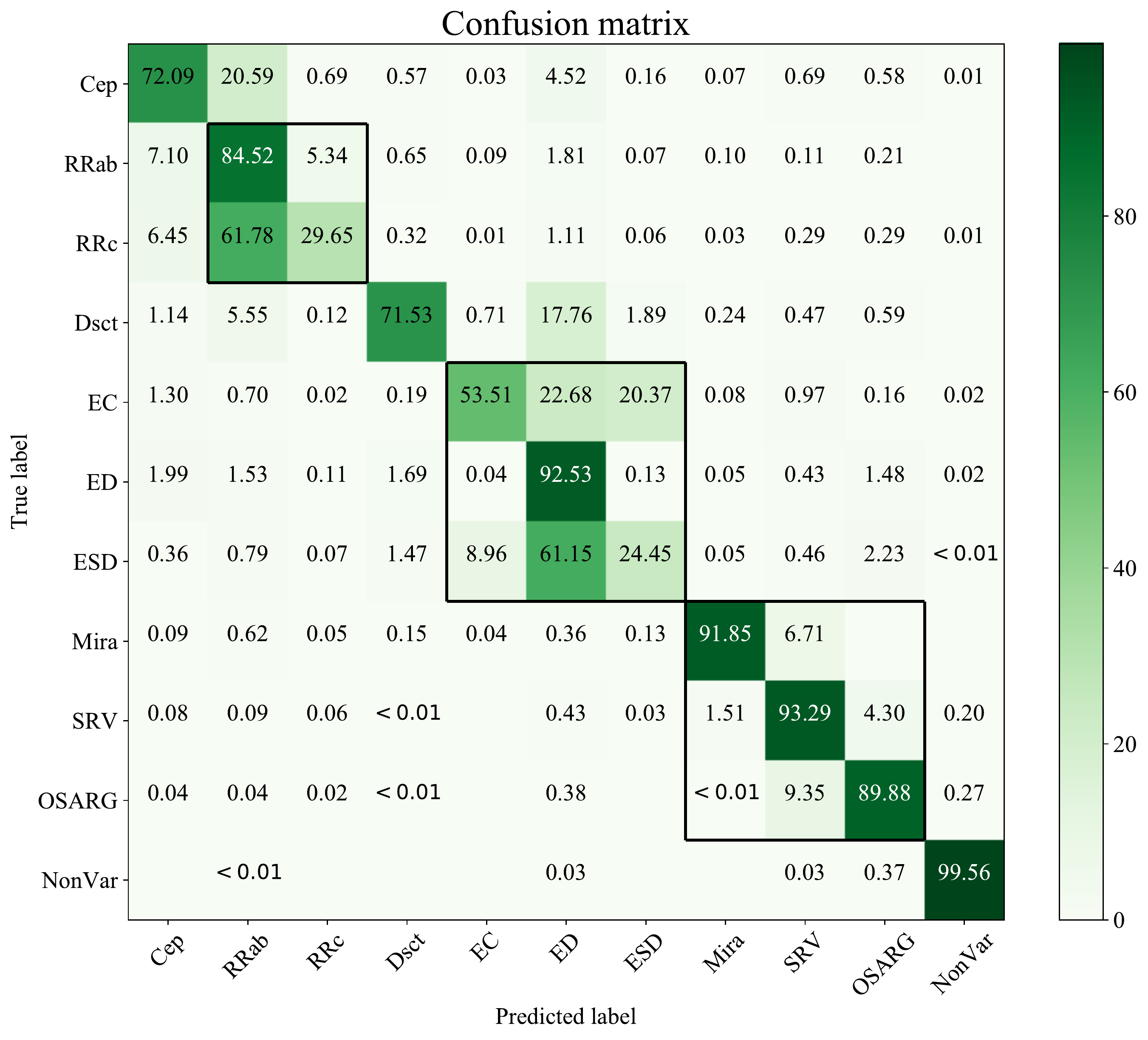}
 \caption{Confusion matrix of the best model for the OGLE-III subclasses. We denote by black squares subclasses that are included within the same ``umbrella class.'' }
 \label{fig:OGLE_CM}
\end{figure*}

Table \ref{tab:classification_ogle} and Figure~\ref{fig:OGLE_CM} show the results for the OGLE-III dataset.

The classification at the class level is comparable with the RF, except for less represented classes, namely Cepheids and $\delta$~Scutis. For eclipsing binaries, LPVs, and RR Lyrae, the majority of the mistakes are done within the class, marked with black lines. For example, \num{61.78}\% of RRc stars get classified as RRab, \num{29.65}\% are correctly classified, and only 8.57\% are classified into other classes. We obtain comparable results with the RF in the well-represented subclasses with at least \num{10000} examples.

\begin{figure*}
 \includegraphics[width=\textwidth,] {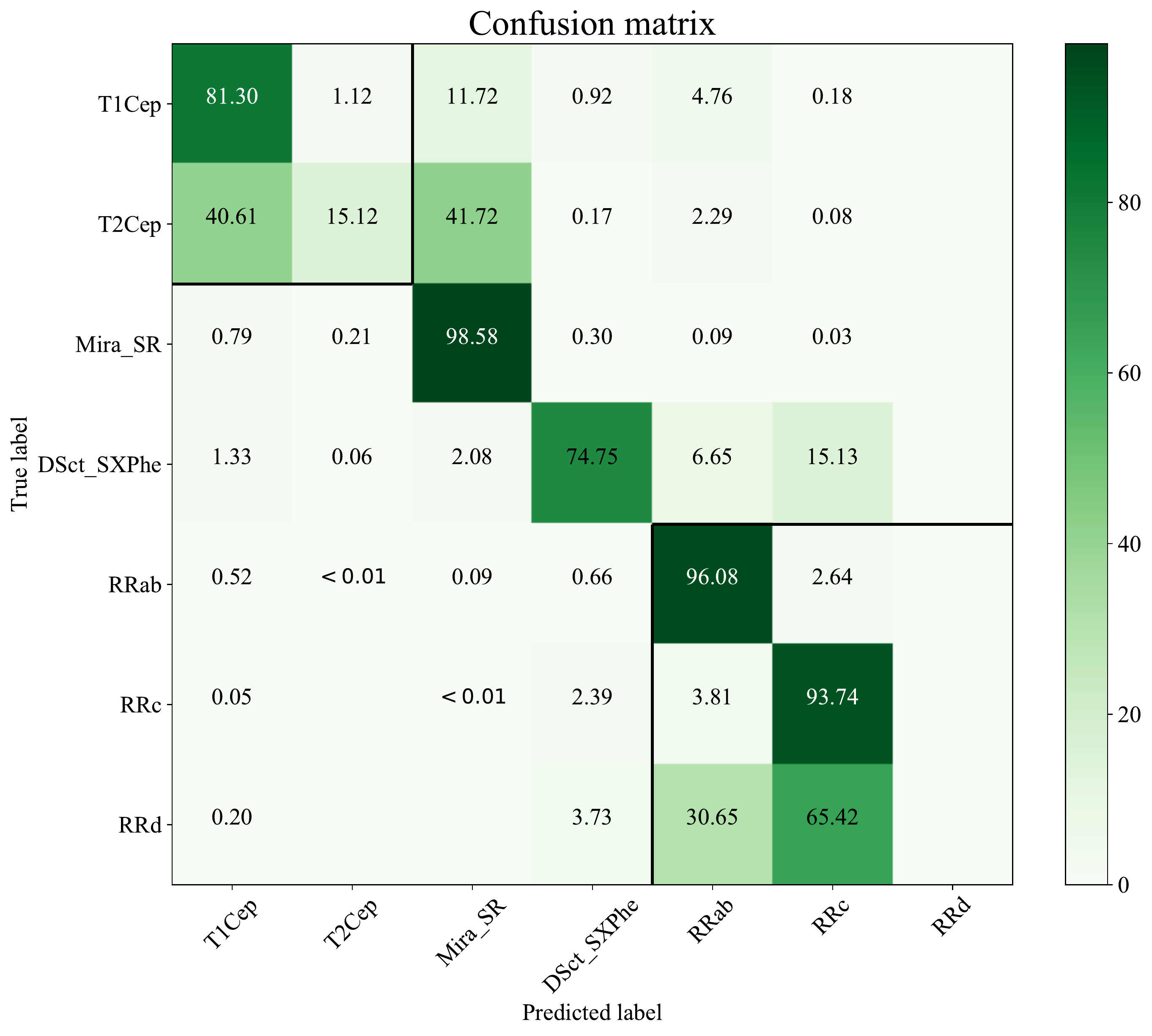}
 \caption{As in Figure~\ref{fig:OGLE_CM}, but for Gaia.}
 \label{fig:GAIA_CM}
\end{figure*}

Similar results are found in the Gaia dataset (Table~\ref{tab:classification_GAIA} and Figure~\ref{fig:GAIA_CM}). At the class level, our model classifies as well as the RF. At subclass level, similar as in OGLE-III, the misclassification is made primarily within the class.
RR Lyrae stars are classified with accuracy greater than \num{90}\%, except the RRd subclass which is classified either as RRab or RRc. The RRd light curves are not long enough to detect multiple periodicities, and have a small sample size. Thus, the network can only recognize one mode of pulsation relative to the other RR Lyrae subtypes. This is also observed in $\delta$~Scuti stars, which are misclassified as RR Lyrae due to the small sample size and comparable pulsation timescales. 

As many as \num{40}\% of type II Cepheids are classified as classical Cepheids, while \num{40}\% are classified as Mira\_SR stars. The misclassification is associated with their pulsation timescales, as roughly 20\% of type II cepheids are RV Tauri stars. Therefore, the estimated periods of this class range from \num{1} to \num{150} days, which intersect the period range of the MIRA\_SR class. Moreover, W Vir, RV Tau, and semi-regular variables can sometimes have morphologically similar light curves \citep{pollard1996rv}, which explains in part the results obtained with our classifier.

The RF results for this class are explained by the poor performance of the period estimation of the FATS package, which assigns periods of less than 100 days for half the MIRA\_SR sample, confusing the classifier with categories such as T2Cep.

The OGLE and Gaia examples are biased because they were selected for being those for which the features produced a good performance. That gives any feature-based classification model an advantage over competitors.
WISE labels were obtained by cross-match and not by their light curve features, thus representing a less biased sample. The RF performance is reduced as features are not well defined in noisier light curves. WISE results need to be addressed in detail, as they provide us with an experiment close to conditions expected at the start of new surveys. 

\begin{figure*}
 \includegraphics[width=\textwidth,]{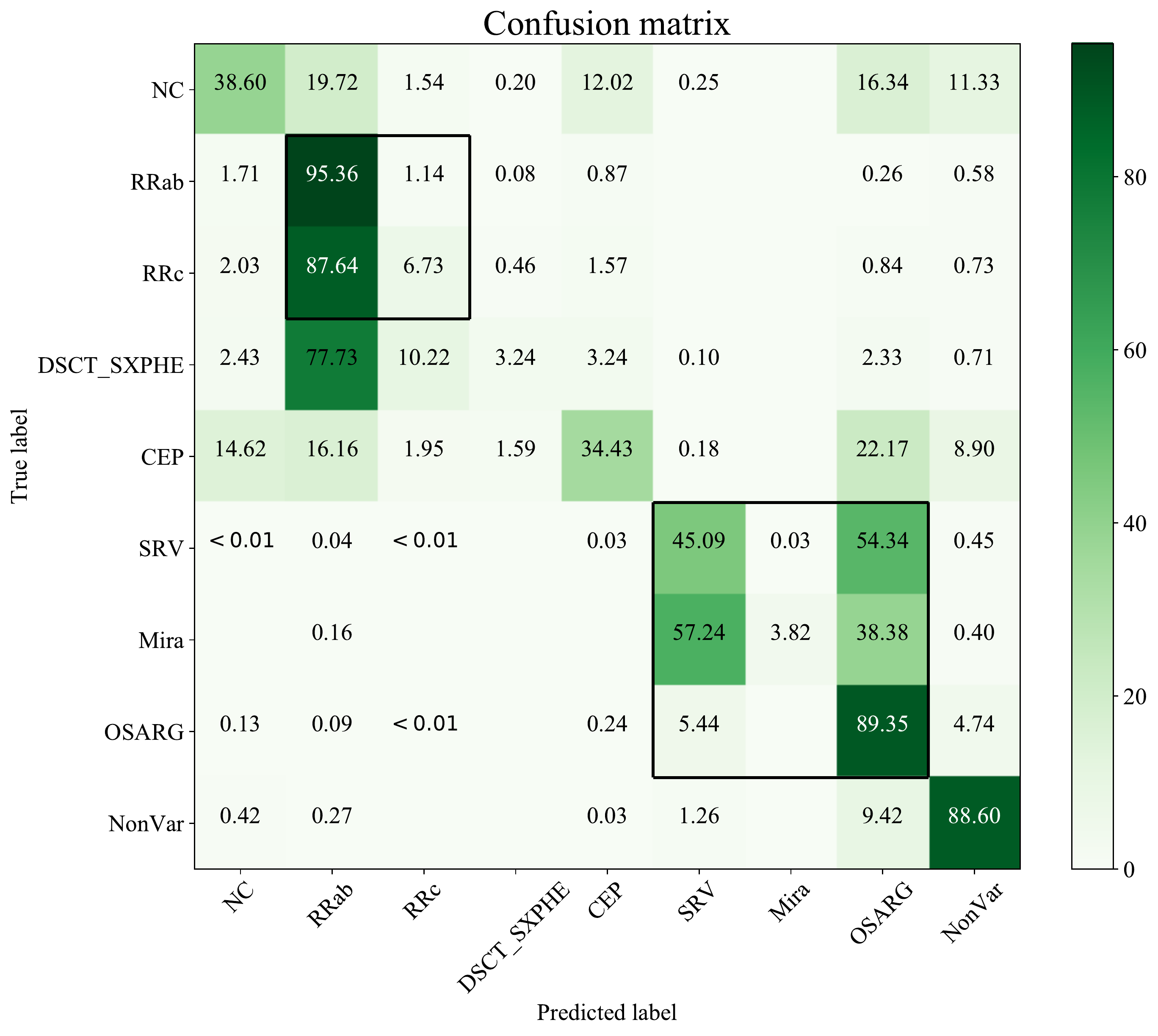}
 \caption{As in Figure~\ref{fig:OGLE_CM}, but for WISE.}
 \label{fig:WISE_CM}
\end{figure*}

The results are shown in Table~\ref{tab:classification_wise} and \ref{fig:WISE_CM}. The classification of RR Lyrae and LPV stars is consistent with the behaviour seen in the other datasets, i.e., the confusion occurs mostly intra-class. As in Gaia, Dsct\_SXPhe are often classified as RR Lyrae. The majority of RRc stars are classified as RRab, and almost all Mira stars are classified as SRVs or OSARGs. As already pointed out previously, the WISE cadence cannot recover complete Mira light curves due to their long periods. Since we are extracting the signal from the light curve directly, our results are worse than with RF.

As expected, OSARGs and non-variable stars are confused the most but in small percentages, because of the small amplitude of OSARGs, and the intrinsic noise of the WISE survey. In other words, the classifier cannot differentiate perfectly whether variability comes from the object or the noise. 

Remarkably, eclipsing binaries and Cepheids are classified by our method with $\sim\num{30}$\% higher accuracy than with the RF. For our model and the RF, both classes are confused the most with NonVar and OSARGs, whose variability is not well defined. Our results are shown in Table~\ref{tab:results_wise_cep_ecl}. The difference in favour of our model is explained by the fact that features cannot reliably characterize light curves that are close to noise. As a result, RF cannot differentiate these variables from non-variable stars.

\begin{table}
    \centering
    \caption{Classification comparison between our method and RF.}
    \label{tab:results_wise_cep_ecl}
    \begin{tabular}{lcc} 
        \hline
        Class & RNN & RF \\
        &  NonVar \& OSARG & NonVar \& OSARG\\
        \hline
        Cep &  \;\;8.90 - 22.17  & 42.00 - 14.54\\
        ECL & 11.33 - 16.34& 45.27 - 16.30 \\
        \hline
    \end{tabular}
\end{table}

The network extracts relevant variability patterns for each subclass. It tends to make mistakes inside each variability class. Even when it misclassifies objects outside their true class, the variability patterns are similar, as in the case of $\delta$~Scuti stars with RR Lyrae, which cover partially overlapping period ranges.
We find that categories with small sample sizes are more prone to be confused with different classes, as seen in the case of non-contact binaries and Cepheids in the WISE dataset.

\begin{figure}
 \includegraphics[width=0.45\textwidth]{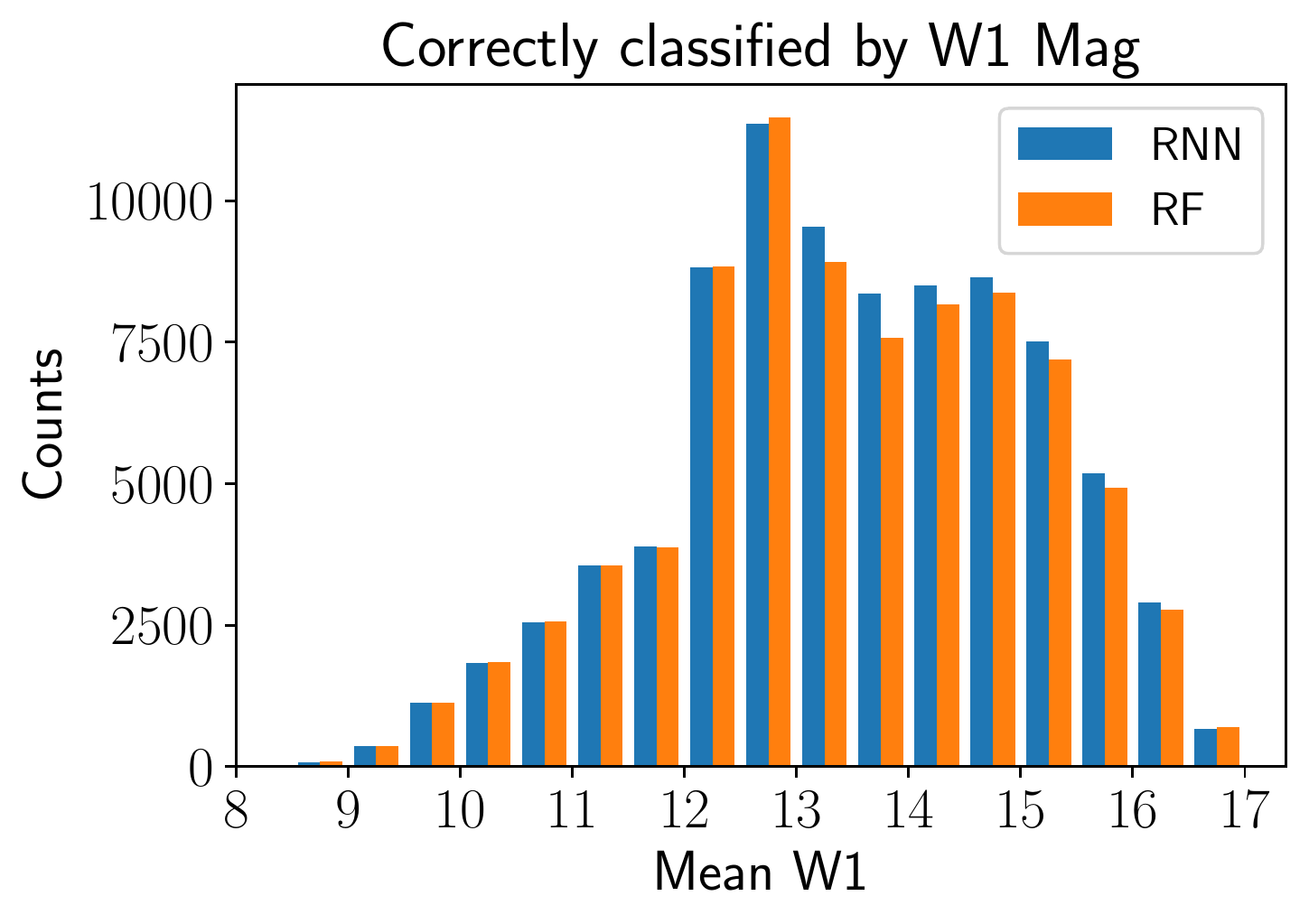}
 \caption{Correctly classified objects by class, as a function of mean $W1$ magnitude (WISE dataset). We consider the same objects for both classifiers. For bright stars, the RF slightly outperforms our model. For fainter objects ($W1 > 13$~mag), our model systematically outperforms the RF, with exception of the last bin.}
 \label{fig:efficiency}
\end{figure}

The fact that we are classifying periodic variables without computing any period at all shows the learning power of neural networks applied in astronomy. Moreover, our results show evidence that neural networks can extract information even in the case of faint objects with very noisy light curves, where traditional features show degraded performance. This effect can be seen in Figure~\ref{fig:efficiency}, where we show the number of correctly classified objects at class level in the WISE dataset, as a function of the mean magnitude. We consider the same objects for both classifiers. The difference in the last bin comes from the greater efficiency of the RF detecting RRab, while our model detected fewer of this type but more of RRc, NC and Cep stars.

\subsection{Computational Run Time}
In Table~\ref{tab:time}, we present the computational runtime of our method and the RF implementation. For the RF, the preprocessing field corresponds to the feature computation time and training for one of the five folds. For the RNN, the preprocessing field corresponds to the transformation into matrix representation and serialization of the light curves, and the training of one fold.

\begin{table}
    \centering
    \caption{Approximate computational runtime in minutes.}
    \label{tab:time}
     \begin{tabular}{lcrrr}
         \toprule
         Survey & Method & Preprocessing & Training & Total \\\midrule
         OGLE-III & RF & 10280 & 9 & 10289\\
          &RNN & 5 & 127 & 132\\\hline
         Gaia & RF & 154 & 4 & 158\\
          &RNN & 2 & 75 & 77\\\hline
         WISE & RF & 271 & 2 & 273\\
          &RNN & 1 & 49 & 50\\\bottomrule
     \end{tabular}
\end{table}

Our method is faster because of the linear scaling in steps and input size, as well as being GPU accelerated. Our network works with minimal preprocessing and an incremental design, which enables it to be trained in a couple of hours at the most. Perhaps most importantly, it can include new data with a marginal added computational cost. 

In comparison, the RF algorithm is trained considerably faster, but the majority of the time is spent in the feature extraction phase, which is done in the CPU. One of the most computationally expensive features is the period, which scales as $O(n\log{n})$ \citep{VanderPlas_2018} with the number of observations $n$. This is the reason why it takes considerably more time to extract features from long, well-populated time series, such as the ones typically found in OGLE-III.

When performing classification, our preprocessing scheme enables us to require only $w-s$ previous observations to update the hidden state and obtain a new label. In contrast, to update some features, FATS requires the entire light curve. This difference reduces the amount of data needed to be transferred from a database, which implies a faster classification with a reduced strain on the system.

Our method shows similar performance as the RF, scales better than the computation of features and enables us to update the classification when new observations are available with a reduced cost compared to features.

\section{Conclusions}\label{section:Conclusion}

In this work, we propose a classification model based on RNNs to perform automatic classification of variable stars. It learns its representation automatically and is designed to work without any pre-computed features.

We test our models in three different surveys: OGLE-III, Gaia DR2, and WISE. We show that a simple neural network architecture can perform on par with state-of-the-art methods such as the RF using FATS features. 

Instead of using magnitudes and times explicitly, our preprocessing stage computes magnitude and time differences for each light curve, and then applies a sliding window sampling scheme, which has low computational cost. This method acts as a normalization scheme, while also serving as a numerical stabilization solution. In a real-time classification scenario, our method avoids unnecessary information requests per object, requiring only a small fraction of the observations, as opposed to traditional feature-based methods. 

Our method scales linearly with the number of points in the light curve. It is designed to be incremental, as the representation of the stars do not need to be recomputed using the entire light curve every time new observations become available. We implement our method to be accelerated by GPU, which means that it can process thousands of objects in parallel, which makes it a viable option to perform classification for the next generation of surveys like the LSST. 

We show that RNNs can learn variability patterns even without the explicit time information, which indicates that the network learns a proxy for the cadence of each survey. 

When performing classification, our model is comparable with the RF. When making mistakes, the majority is made inside a given ``umbrella class'': for example, RRc stars may be misclassified as RRab stars. Our data does not sample well the period space, being the RR Lyraes and LPVs the most numerous. As such, objects with intermediate periods tend to be missclassified among those two extremes. As in the case of type 2 Cepheids being misclassified as LPVs.

Against the RF, for a less biased dataset such as WISE, the RNN can separate Cepheids from the intrinsic noise of non-variable stars with greater efficiency. Additionally, we show that our classifier has better performance at higher mean magnitudes.

Computing features is still a crucial part of the study of variable stars. We propose a mixed model of classification, where we leverage the advantages of our feature-less classification to reduce the number of candidates for a specific category. Then, features can be computed by taking into account the class of the object.
The advantage is that feature extraction can be optimized. For example, the period search can be optimized by using different search strategies for each class like RR Lyrae stars and LPVs, which have periods which differ by orders of magnitude. Optimizing the feature extraction and computing just the necessary periods could speed up the classification and discovery of new variables.

These learned parameters are optimized to reduce the classification loss, and as such, do not represent any physical quantity in a direct form. We recognize the importance of understanding our model, which we leave as future work. Specifically, we aim to study why eclipsing binaries and Cepheids show better performance than other classes.

Our future work aims to extend our method to include multi-band observations, additional physical information and to leverage the class hierarchy, to improve the performance in streaming classification. Another objective is to include the less represented classes as well as gravitational microlensing and other transient and non-periodic variability phenomena. 

Our methods show potential to be scalable and fast enough to be applied to a stream of data which is expected to arrive with the LSST. Our implementation, including Python code and datasets, is available online.  \footnote{\url{https://github.com/iebecker/Scalable_RNN}}

\section*{Acknowledgments} \label{section:Acknowledgments}

We acknowledge the support from CONICYT-Chile, through CONICYT-PFCHA/Doctorado Nacional/2018-21181990 and FONDECYT Regular projects number 1180054 and 1171273.
Additional support for this project is provided by by Proyecto Basal AFB-170002 and by the Ministry for the Economy, Development, and Tourism's Millennium Science Initiative through grant IC\,120009, awarded to the Millennium Institute of Astrophysics (MAS); by Proyecto Basal PFB-06/2007; and by FONDECYT grant \#1171273. FN is grateful for financial support by Proyecto Gemini CONICYT grants \#32130013, \#32140036 and VRI.

This work has made use of data from the European Space Agency (ESA) mission \textit{Gaia}, processed by the \textit{Gaia} Data Processing and Analysis Consortium (DPAC). Funding for the DPAC has been provided by national institutions, in particular the institutions participating in the \textit{Gaia} Multilateral Agreement.
 


\bibliographystyle{mnras}
\bibliography{biblio}


%
%


\bsp    
\label{lastpage}
\end{document}